\documentclass{aa}
\usepackage[varg]{txfonts}
\usepackage{natbib}
\usepackage{comment}

\newcommand{\fei}{Fe\,{\sc i}}
\newcommand{\feii}{Fe\,{\sc ii}}
\newcommand{\mgi}{Mg\,{\sc i}}
\newcommand{\baii}{Ba\,{\sc ii}}
\makeatletter
\newcommand\footnoteref[1]{\protected@xdef\@thefnmark{\ref{#1}}\@footnotemark}
\makeatother

\begin{document}

\title{The $Gaia$-ESO Survey: the inner disk intermediate-age open  cluster NGC 6802\thanks{Tables 3 to 6 are only available in electronic form at the CDS via anonymous ftp to cdsarc.u-strasbg.fr (130.79.128.5) or via http://cdsweb.u-strasbg.fr/cgi-bin/qcat?J/A+A/.} }


\author{B. Tang\inst{1}
  \and D. Geisler\inst{1}
  \and E. Friel\inst{2}
  \and S. Villanova\inst{1}
  \and R.~Smiljanic\inst{3}  
  \and A.~R. Casey\inst{4}
  \and S. Randich\inst{5}  
  \and L.~Magrini\inst{5}    
  \and I. San Roman\inst{6}
  \and C. Mu\~{n}oz\inst{1,7}
  \and R.~E. Cohen\inst{1}
  \and F. Mauro\inst{1}
  \and A. Bragaglia\inst{8}
  \and P. Donati\inst{8}  
  \and G.~Tautvai\v{s}ien\.{e}\inst{9}
  \and A. Drazdauskas\inst{9}
  \and R. \v{Z}enovien\.{e}\inst{9}
  \and O.~Snaith\inst{10}    
  \and S. Sousa\inst{11} 
  \and V. Adibekyan\inst{11}  
  \and M. T. Costado\inst{12}
  \and S. Blanco-Cuaresma\inst{13}
  \and F. Jim\'{e}nez-Esteban\inst{14,15}
  \and G. Carraro\inst{16}
  \and T. Zwitter\inst{17}
  \and P. Fran\c{c}ois\inst{18} 
  \and P. Jofr\`e\inst{4,19}
  \and R. Sordo\inst{20}
  \and G. Gilmore\inst{4} 
  \and E. Flaccomio\inst{21}
  \and S. Koposov\inst{4} 
  \and A.~J. Korn\inst{22}
  \and A.~C. Lanzafame\inst{23,24}  
  \and E. Pancino\inst{5,25}         
  \and A. Bayo\inst{26}
  \and F. Damiani\inst{21}
  \and E. Franciosini\inst{5} 
  \and A. Hourihane\inst{4}    
  \and C. Lardo\inst{27}
  \and J. Lewis \inst{4} 
  \and L. Monaco\inst{28}   
  \and L. Morbidelli\inst{5}
  \and L. Prisinzano\inst{21}  
  \and G. Sacco\inst{5}
  \and C.~C. Worley\inst{4}
  \and S. Zaggia\inst{20}
}


\institute{Departamento de Astronom\'ia, Casilla 160-C, Universidad de Concepci\'on, Concepci\'on, Chile\\
                 email: btang@astro-udec.cl
  \and Department of Astronomy, Indiana University, Bloomington, IN, USA
  \and Nicolaus Copernicus Astronomical Center, Polish Academy of Sciences, Bartycka 18, 00-716, Warsaw, Poland 
  \and Institute of Astronomy, University of Cambridge, Madingley Road, Cambridge CB3 0HA, UK
  \and INAF $-$ Osservatorio Astrofisico di Arcetri, Largo E. Fermi, 5, 50125, Florence, Italy     
  \and Centro de Estudios de F\'isica del Cosmos de Arag\'on (CEFCA), Plaza San Juan 1, 44001 Teruel, Spain
  \and European Southern Observatory, Casilla 19001, Santiago, Chile 
  \and INAF $-$ Osservatorio Astronomico di Bologna, Via Ranzani, 1, 40127, Bologna, Italy
  \and Institute of Theoretical Physics and Astronomy, Vilnius University, Sauletekio av. 3, LT-10222 Vilnius, Lithuania
  \and School of Physics, Korea Institute for Advanced Study, 85 Hoegiro, Dongdaemun-gu, Seoul 02455, Republic of Korea
  \and Instituto de Astrof\'{i}sica e Ci\^{e}ncias do Espa\c{c}o, Universidade do Porto, CAUP, Rua das Estrelas, 4150-762 Porto, Portugal
  \and Instituto de Astrof\'{i}sica de Andaluc\'{i}a-CSIC, Apdo. 3004, 18080 Granada, Spain
  \and Observatoire de Gen\`eve, Universit\'e de Gen\`eve, CH-1290 Versoix, Switzerland
  \and Centro de Astrobiolog\'{i}a (INTA-CSIC), Dpto. de Astrof\'{i}sica, PO Box 78, 28691 Villanueva de la Ca\~{n}ada, Madrid, Spain
   \and Suffolk University, Madrid Campus, C/ Valle de la Vi\~{n}a 3, 28003, Madrid, Spain
   \and Department of Physics and Astronomy, University of Padova, Via Marzolo, 8, 35131 Padova, Italy
   \and University of Ljubljana, Faculty of Mathematics and Physics, Ljubljana, Slovenia
   \and GEPI, Observatoire de Paris, CNRS, Universit\'e Paris Diderot, 5
Place Jules Janssen, 92190 Meudon, France
  \and N\'ucleo de Astronom\'ia, Facultad de Ingenier\'ia, Universidad Diego Portales, Av. Ejercito 441, Santiago, Chile 
  \and INAF $-$ Padova Observatory, Vicolo dell’Osservatorio 5, 35122, Padova, Italy  
  \and INAF $-$ Oss. Astronomico di Palermo, Piazza del Parlamento 1, 90134 Palermo, Italy  
  \and Department of Physics and Astronomy, Uppsala University, Box 516, SE-751 20 Uppsala, Sweden  
  \and Universit\`a di Catania, Dipartimento di Fisica e Astronomia, Sezione Astrofisica, Via S. Sofia 78, I-95123 Catania, Italy
  \and INAF $-$ Osservatorio Astrofisico di Catania, Via S. Sofia 78, I-95123 Catania, Italy    
  \and ASI Science Data Center, Via del Politecnico SNC, 00133 Roma, Italy
  \and Instituto de F\'isica y Astronom\'ia, Universidad de Valpara\'iso, Chile 
  \and Astrophysics Research Institute, Liverpool John Moores University, 146 Brownlow Hill, Liverpool L3 5RF, United Kingdom
  \and Departamento de Ciencias Fisicas, Universidad Andres Bello, Republica 220, Santiago, Chile
}

\date{Received   / Accepted  }

\abstract {Milky Way open clusters are very diverse in terms of age, chemical composition, and kinematic properties. Intermediate-age and old open clusters are less common, and it is even harder to find them inside the solar Galactocentric radius, due to the high mortality rate and strong extinction inside this region. NGC 6802 is one of the inner disk open clusters (IOCs) observed by the $Gaia$-ESO survey (GES). This cluster is an important target for calibrating the abundances derived in the survey due to the kinematic and chemical homogeneity of the members in open clusters. Using the measurements from $Gaia$-ESO internal data release 4 (iDR4), we identify 95 main-sequence dwarfs as cluster members from the GIRAFFE target list, and eight giants as cluster members from the UVES target list. The dwarf cluster members have a median radial velocity of $13.6\pm1.9$ km s$^{-1}$, while the giant cluster members have a median radial velocity of $12.0\pm0.9$ km s$^{-1}$ and a median [Fe/H] of $0.10\pm0.02$ dex. The color-magnitude diagram of these cluster members suggests an age of $0.9\pm0.1$ Gyr, with $(m-M)_0=11.4$ and $E(B-V)=0.86$. We perform the first detailed chemical abundance analysis of NGC 6802, including 27 elemental species. To gain a more general picture about IOCs, the measurements of NGC 6802 are compared with those of other IOCs previously studied by GES, that is, NGC 4815, Trumpler 20, NGC 6705, and Berkeley 81. NGC 6802 shows similar C, N, Na, and Al abundances as other IOCs. These elements are compared with nucleosynthetic models as a function of cluster turn-off mass.
The $\alpha$, iron-peak, and neutron-capture elements are also explored in a self-consistent way.
}
  
\maketitle 
\section{Introduction}
\label{sect:intro}
In recent years, large Galactic spectroscopic surveys with high resolution multi-object spectrographs, such as the $Gaia$-ESO survey (GES; \citealt{Gilmore2012, Randich2013}), the Apache Point Observatory Galactic Evolution Experiment (APOGEE) survey \citep{Majewski2015}, 
the RAdial Velocity Experiment (RAVE) survey \citep{Siebert2011}, and the Galactic Archaeology with HERMES (GALAH) survey \citep{DeSilva2015}, have greatly improved our knowledge about the chemical and kinematic properties of the Milky Way by providing high quality and homogeneously reduced spectra for more than $\sim$10$^5$ stars.

Open clusters are ideal laboratories to trace Galactic evolution. Their homogeneous chemical compositions are frequently used to calibrate the chemical abundances derived from large Galactic spectroscopic surveys \citep[e.g.,][]{Smiljanic2014, Meszaros2013}.
However, it is not easy for open clusters to survive in the energetic part of the Galaxy -- the disk \citep[e.g.,][]{Freeman2002,Moyano2013} -- where open clusters generally last for only a few hundred million years. In particular, there is enhanced gravitational interaction in the inner disk, interior to the Sun \citep{Friel1995}. Therefore, any surviving intermediate-age or old open clusters are particularly valuable resources for studying the disk. Due to the strong extinction inside the solar Galactocentric radius\footnote{$\sim$8 kpc \citep{Francis2014}.}, observations of inner disk open clusters (hereafter IOCs) are particularly challenging. GES has targeted 12 IOCs up through internal data release 4 (iDR4). Four of them have been analyzed using previous data releases: NGC 4815 \citep{Friel2014}, Trumpler 20 \citep{Donati2014}, NGC 6705 \citep{CantatGaudin2014}, and Berkeley 81 \citep{Magrini2015}. Trumpler 23 is also under analysis (Overbeek et al., accepted). In this paper, we will closely inspect the chemical abundances of another IOC, NGC 6802. Readers are referred to \citet{Jacobson2016} for a complete list of the 12 intermediate-age or old IOCs observed in GES. These IOCs are located between 5$-$8 kpc away from the Galactic center (R$_{\rm GC}$), and within 150 pc from the mid-plane. Five of them show ages older or roughly equal to 1 Gyr, including NGC 6802.

The IOC of this work, NGC 6802 ($l=55.32^{\circ}$, $b=0.92^{\circ}$, $\alpha_{\rm J2000} =  292.6500^{\circ}$, $\delta_{\rm J2000} = -20.2594^{\circ}$), has been studied previously with photometry and spectroscopy. \citet{Sirbaugh1995} published an abstract about the basic properties of NGC 6802 using the $BVRI$ bands. They found a best-fit [Fe/H] of $-0.45$, with a distance modulus ($(m-M)_0$) of 10.78, a reddening (E($B-V$)) of 0.94, and an age of 1.0 Gyr. Later, \citet[][JH11]{Janes2011} estimated $\log(\rm{age}) = 8.98 \pm 0.07$, $(m-M)_0=11.3$, E($B-V$)$=0.84$, and $R_{\rm {GC}} =7.4$ kpc, also with $BVRI$ photometry. These two studies agree well with each other in terms of age, $(m-M)_0$, and E($B-V$). 
At the same time, JH11 also estimated the photometric membership probability. First, the cluster center and radius were estimated by fitting Gaussian plus quadratic background functions to the marginal distributions along both RA (right ascension) and DEC (declination) directions. 
They selected the value equal to two-thirds of the average FWHM (full width at half maximum) values of the RA and DEC marginal distributions as the cluster radius. The density of stars is significantly higher inside the JH11 radius than in the surrounding field regions.
The cluster radius of NGC 6802 was found to be $100''$. All of the cluster members in their works are located inside this radius. Then the membership probability was defined as one minus the ratio of the field star density divided by the cluster star density at the position of the target star in the color magnitude diagram (CMD). Stars with photometric membership probability greater than 50\% are assumed to be possible cluster members.
Subsequently, \citet{Hayes2014} observed stars in the field of NGC 6802 using the Wisconsin$-$Indiana$-$Yale$-$NOAO (WIYN) 3.5 m telescope,  Bench Spectrograph (0.158 \AA~pixel$^{-1}$, 6082$-$6397 \AA). A total of 25 cluster members were identified with a mean radial velocity (RV) of $12.4 \pm 2.8$ km s$^{-1}$. These data were taken as inputs for the GES target selection, and to derive velocities to help with membership determination for evolved stars.

In this paper, we take advantage of the GES iDR4 to analyze the chemical and kinematic properties of NGC 6802 (Sect. \ref{sect:data}). We constrain age, reddening, and distance of this open cluster with the cluster members identified from GIRAFFE and UVES (Ultraviolet and Visual Echelle Spectrograph) measurements.
We combine our analysis of NGC 6802 with those of four previously published GES IOCs: NGC 4815, Trumpler 20 (Tr 20), NGC 6705, and Berkeley 81 (Be 81) to discuss the chemical abundances seen in these similar stellar systems (Sect. \ref{sect:re}). A larger set of neutron-capture elements recently available now allow the discussion of different types of neutron-capture processes (Sect. \ref{sect:nc}). In Sect. \ref{sect:dis}, we discuss possible s-process enhancements in IOCs compared to the control samples. 
Finally, a brief summary of the results is given in Sect. \ref{sect:con}.

\begin{figure}
\includegraphics [width=0.5\textwidth]{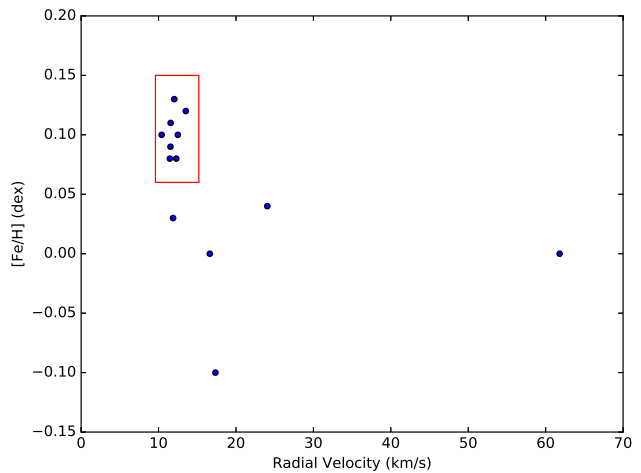} 
\caption{ [Fe/H] as a function of radial velocity (RV). All stars observed with UVES (Ultraviolet and Visual Echelle Spectrograph) are plotted. Selected cluster members are those inside the rectangle. The rectangle indicates the region that satisfies $9.6<{\rm RV}<15.2$ km s$^{-1}$ and $0.06<[$Fe/H$]<0.15$ dex.}\label{fig:vfe}
\end{figure}

\begin{figure}
\centering
\includegraphics [width=0.5\textwidth]{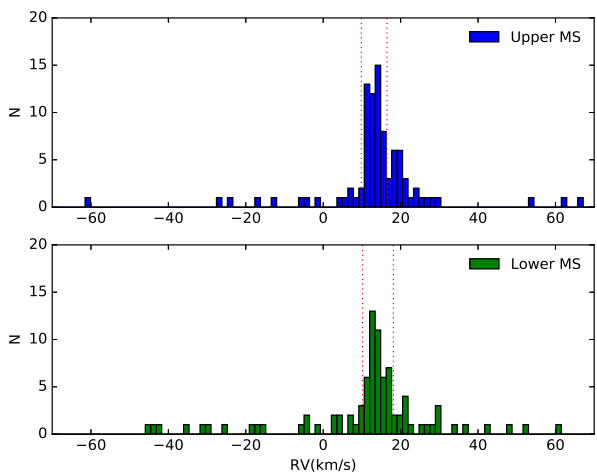} 
\caption{ Radial velocity distributions for upper main sequence stars (upper panel) and lower main sequence stars (lower panel). The red dotted lines of each panel indicate the lower and upper limits of the 2-sigma cut used to limit cluster membership. See text for more details. }\label{fig:vdist}
\end{figure}

\section{Target selection and observation}
\label{sect:obs}

The $Gaia$-ESO survey \citep{Gilmore2012, Randich2013} utilizes the FLAMES (Fiber Large Array Multi-Element Spectrograph) of the VLT (Very Large Telescope) to study the kinematic and chemical evolution of our Milky Way.
Up to 10$^5$ stars in the Galaxy are being obtained by the medium-resolution ($R\sim20 \,\, 000$) multi-object spectrograph GIRAFFE. The UVES instrument will also observe up to 5000 stars with  high spectral resolution ($R \approx 47 \,\,000$).

To select possible cluster members of NGC 6802, we considered the cluster center (RA$=$19:30:35, Dec$=+$20:15:42) and radius (5 arcmin) from WEBDA.\footnote{http://www.univie.ac.at/webda/.} For main sequence (MS) stars, we selected from the complete photometric catalog of JH11 all stars within the WEBDA radius (5 arcmin). For the stars in the red clump, we selected a more stringent criterion in radius, choosing stars within the JH11 cluster radius ($100''$).

\subsection{MS}
To define the MS locus, we employed the Padova isochrone with the same parameters given in JH11. To select stars on the MS we used the isochrone as a zero-point and we selected all stars within an envelope of $V-I=0.1$ mag to the left and right sides of the isochrone. MS stars were divided into A type stars, essentially brighter than $V=17$, and G type stars, in the magnitude range $V=17-19$. We used GIRAFFE to observe G type and A type stars in three different configurations, namely:
\begin{itemize}
\item bright A type stars: observed with HR09B grating setup (5143$- $5356 \AA~and $R=25~900$) for two hours;
\item faint A type stars and bright G type stars: observed with HR09B grating setup for three hours;
\item G type stars: observed with HR15N grating setup (6470$- $6790 \AA~and $R=17~000$) for five hours.
\end{itemize}
Two different grating setups were chosen partially due to the available spectral lines and the blackbody radiation of A type and G type stars. We note that there are some overlaps between the three configurations.

\subsection{Red clump}
We selected 21 stars located within the JH11 cluster radius ($100''$), and with colors and magnitudes consistent with the red clump of the Padova isochrone. Among them, we selected 14 stars with the radial velocity information given by the WIYN observations (Sect.  \ref{sect:intro}).  These stars were divided into two groups, where each group of seven stars was observed by UVES (4700$- $6840 \AA~with a gap of $\sim50$ \AA~in the center, $R=47~000$), simultaneously with one GIRAFEE grating setup. The total exposure time for each UVES$+$GIRAFFE grating setup is five hours. However, one red clump star was observed twice in the actual observation, leaving another candidate red clump star unobserved, thus only 13 stars were given in iDR4.

\section{Cluster members identified by radial velocity}
\label{sect:data}

Analysis of the GIRAFFE FGK star atmospheric parameters
and abundances (using solar abundances by \citealt{Grevesse2007}, hereafter G07) is handled by WG (work group) 10 (Recio-Blanco et al., in prep.), while UVES FGK star parameters and abundances are derived by WG 11. The spectral reduction and radial velocity derivation for UVES spectra are described in \citet{sacco2014}. 
In WG 11, the stellar abundances are derived by individual nodes using different techniques (e.g., equivalent width, spectral synthesis). An error model is constructed for each node based on their performance on the well-studied benchmark stars. Covariances between nodes are accounted for in the recommended parameters (Casey et al. 2017, in preparation). For a detailed description of the node analysis techniques and homogenization of DR3 data, see \citet{Smiljanic2014}. DR3 is the latest public data, which can be accessed through http://ges.roe.ac.uk/.
The atmospheric parameters and element abundances in this paper come from the recommended abundance table of iDR4. Compared to the former data releases, the number of available chemical species in iDR4 has greatly increased, especially for the neutron-capture elements. 

\subsection{UVES members}

A clustering of stars around RV$\sim12$ km s$^{-1}$ and [Fe/H$]\sim0.1$ dex is clearly spotted in Figure \ref{fig:vfe}.
It shows that eight stars are inside the red rectangle, which has dimensions of 0.09 dex in metallicity, and 5.6 km s$^{-1}$ in radial velocity. The rest of the stars have lower iron abundances, and several also have disparate velocities. The eight cluster members show a median RV of $12.0\pm0.9$ km s$^{-1}$ and a median [Fe/H] of $0.10\pm0.02$ dex. We will consider the detailed chemical abundances of these eight stars in the next section. Information on the UVES cluster members is listed in Table \ref{tab1}. The non-members observed by UVES are listed in Table \ref{tabnon}.  We note that the red clump star that was observed twice is Star N1 in Table \ref{tabnon}.

\begin{figure*}
\centering
\includegraphics [width=\textwidth]{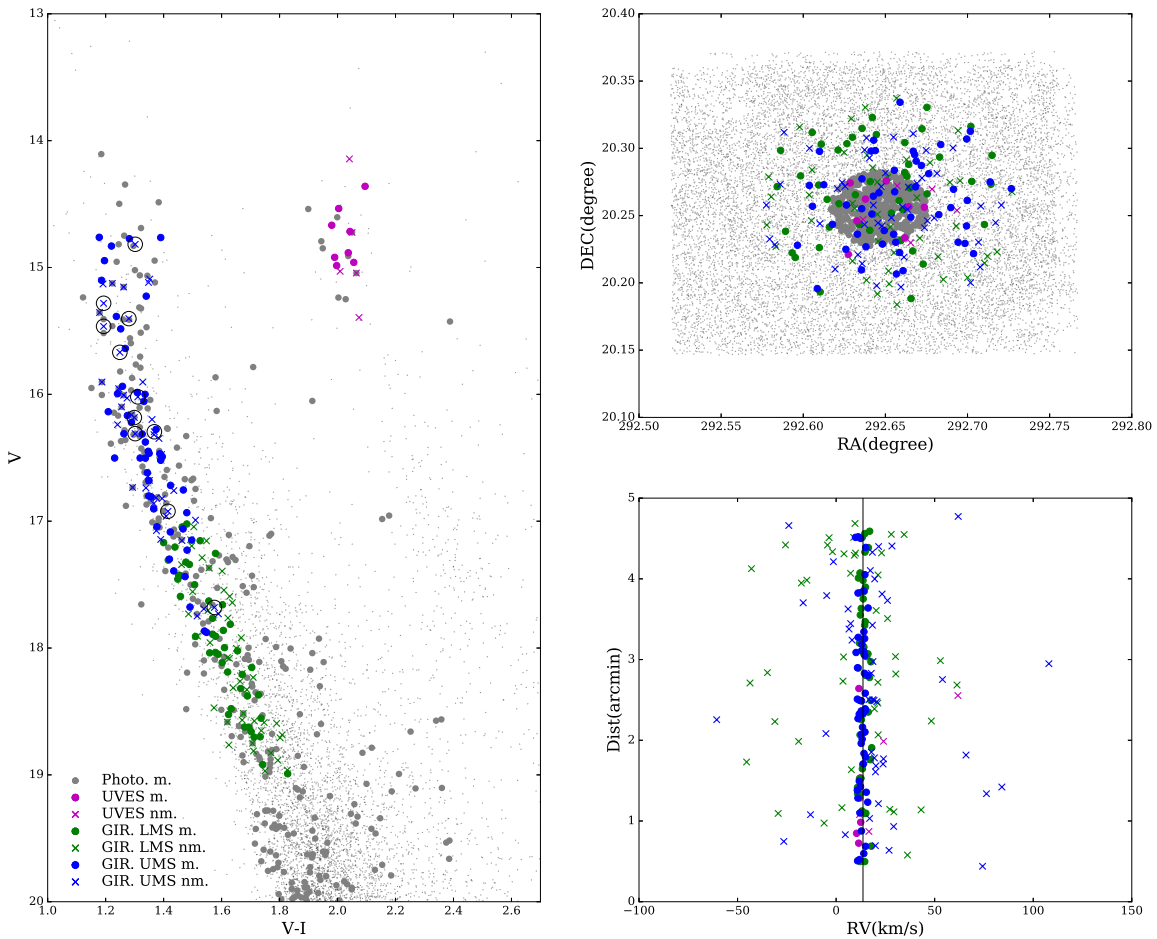} 
\caption{Left: Color-magnitude diagram of NGC 6802 stars. The background dots are photometric data from \citet{Janes2011}. Stars with photometric membership probability greater than 50\% are shown as gray filled circles. We also plot the UVES cluster members (magenta filled circles),  UVES non-members (magenta crosses), GIRAFFE LMS (lower main sequence) cluster members (green filled circles), GIRAFFE LMS non-members (green crosses), GIRAFFE UMS (upper main sequence) cluster members (blue filled circles), and GIRAFFE UMS non-members (blue crosses). Empty circles indicate UMS stars with RV between 18 and 21 km s$^{-1}$. The symbol meanings are also applicable to the other panels. Top~right: Locations of stars observed by \citet{Janes2011}, UVES, and GIRAFFE. Bottom~right: Distance from the cluster center as a function of radial velocity. The median radial velocity of the GIRAFFE members is marked as a solid line.}\label{fig:giruv}
\end{figure*}

\subsection{GIRAFFE members}

We found 184 target stars observed by GIRAFFE in the field of NGC 6802. One star which shows obviously much lower radial velocity ($-345.63$ km s$^{-1}$) is excluded from the discussion below. The RV distribution of the other 183 stars peaks at RV$\sim 14$ km s$^{-1}$ (Figure \ref{fig:vdist}). As shown in Sect.  \ref{sect:obs}, these stars were observed with HR15N  and/or HR9B setups.  When both HR15N and HR9B observations are available for the same star, the former setup is chosen for analysis by GES, due to longer exposed time. The radial velocities of 88 stars observed by the HR15N setup and 95 stars observed by the HR9B setup are given in iDR4. Since the stars observed by the former setup locate in the lower part of the MS, we call these stars LMS (a more detailed description of the photometry is given below). For a similar reason, the stars observed with the latter setup are called upper MS (UMS, Figure \ref{fig:giruv}). On average, LMS stars show smaller RV errors (0.82 km s$^{-1}$) than that of UMS (2.2 km s$^{-1}$), mainly because it is easier to measure RV in cooler stars. The RV distribution of LMS is more symmetric, while the RV distribution of UMS is asymmetric, with a possible second peak around 20 km s$^{-1}$ (Figure \ref{fig:vdist}). For these reasons, we treat LMS and UMS separately in this paper. 
We apply 2-sigma clipping to the RV distribution functions of both LMS and UMS. After the clipping, 45 stars are left in LMS (Table \ref{tablms}). The lower and upper limits of the clipping are 10.2 and 18.1 km s$^{-1}$, respectively (red dotted lines in the lower panel of Figure \ref{fig:vdist}). These 45 LMS cluster members show a median  RV of $14.1\pm2.0$ km s$^{-1}$.
At the same time, 50 stars are left in UMS after clipping (Table \ref{tabums}). The stars in the second peak are excluded during the clipping, and the RV errors of these stars are statistically larger than other cluster members (Sect.  \ref{sect:iso}). The lower and upper limits of the clipping are 9.8 and 16.5 km s$^{-1}$, respectively (red dotted lines in the upper panel of Figure \ref{fig:vdist}). These 50 UMS cluster members show a median RV of $13.2\pm1.7$ km s$^{-1}$. The non-members observed by GIRAFFE are listed in Tables \ref{tab:lmsno} and \ref{tab:umsno}. In total, 95 member stars are identified by GIRAFFE spectra, with a median RV of $13.6\pm1.9$ km s$^{-1}$. The median measurement error for the GIRAFFE member stars is 1.1 km s$^{-1}$. In contrast, the UVES members show a median RV of 12.0 km s$^{-1}$, with a dispersion of 0.9  km s$^{-1}$ and a measurement error of 0.57 km s$^{-1}$. The RVs obtained from UVES and GIRAFFE samples agree within the uncertainty ranges of both dispersion and measurement error. 

\subsection{Estimated age from isochrone fitting}
\label{sect:iso}

To construct the CMD, we retrieve the iDR4 photometric data of our cluster members\footnote{From the ``target'' table of iDR4.}, which are in fact drawn from the photometric data of JH11. All of the stars observed by JH11 in the field of NGC 6802 are plotted as small dots in Figure \ref{fig:giruv}. Stars with photometric membership probability greater than 50\% are shown as gray filled circles in Figure \ref{fig:giruv}. Cluster members confirmed by GES RV are labeled with filled circles of different colors for each sample. We note that 30 giants and dwarfs with photometric membership probability greater than 50\% are identified as spectroscopic members. The remaining 73 RV-confirmed members are mainly located outside the cluster radius estimated by JH11.  This makes sense, because JH11 specifically pointed out `these radii, ..., represent effective sizes of the clusters, within which the density of stars is significantly higher than in the surrounding field regions.'
The actual cluster tidal radius must be larger than $100''$ for NGC 6802. It may reach $276''$, which is the largest distance to the cluster center for the RV-confirmed members (Figure \ref{fig:giruv}). There may be concern about losing stars in open clusters \citep{Davenport2010,Dalessandro2015}, which makes it difficult to precisely calculate structural quantities like tidal radius that carry an assumption of dynamical equilibrium, but extra information, for example, proper motion, may be needed to confirm this.


We also pick out UMS stars with RV between 18 and 21 km s$^{-1}$, where a possible second peak in UMS is spotted, and label these stars with empty circles (Figure \ref{fig:giruv}). Table \ref{tab:umsno} gives more details.  These stars tend to have lower accuracy in RV (the median RV error is 2.9 km s$^{-1}$) due to broader lines in hot stars. However, these stars have been excluded by the 2-sigma clipping, so they will not affect our discussions below. Interestingly, a possible second peak is also spotted by \citet{Prisinzano2016} in $\gamma$ Velorum cluster MS dwarfs. They speculated that this structure may be related to the large uncertainties in the RV measurements of fast rotators.  

\begin{figure}

\includegraphics [width=0.5\textwidth]{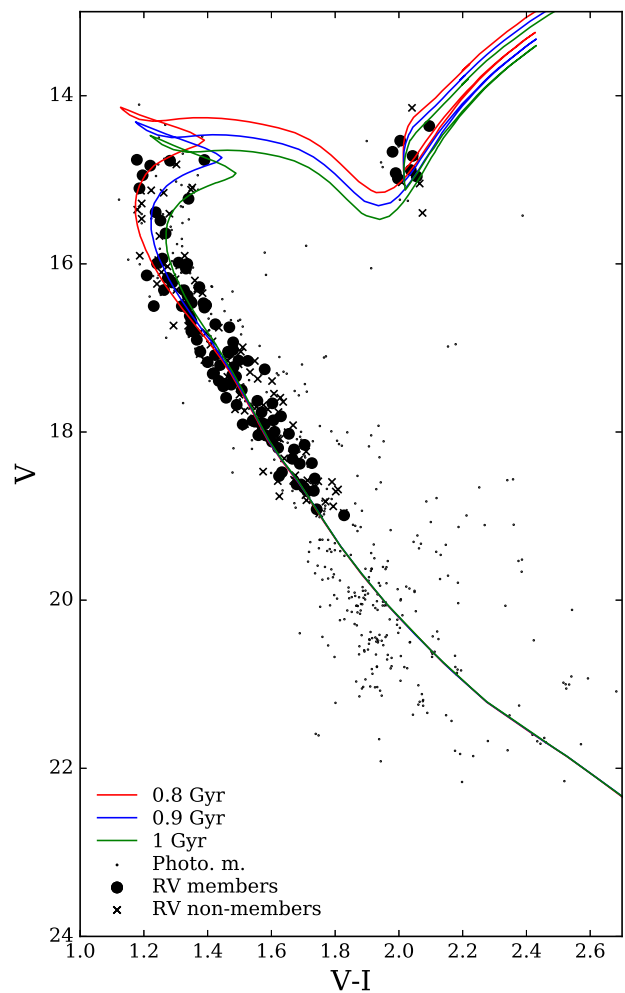} 
\caption{Color-magnitude diagram of NGC 6802 stars. Stars with photometric membership probability greater than 50\% are shown as small black dots. Cluster members (black filled circles) and non-members (black crosses) identified by spectral RV are also plotted. Red, blue, and green solid lines represent the PARSEC isochrones with ages of 0.8, 0.9, and 1.0 Gyr, respectively. For all three isochrones, [Fe/H$]=0.1$ dex, E$(B-V)=0.86$, and $(m-M)_0=11.4$. 
}\label{fig:cmd}
\end{figure}

\begin{figure}
\includegraphics [width=0.5\textwidth]{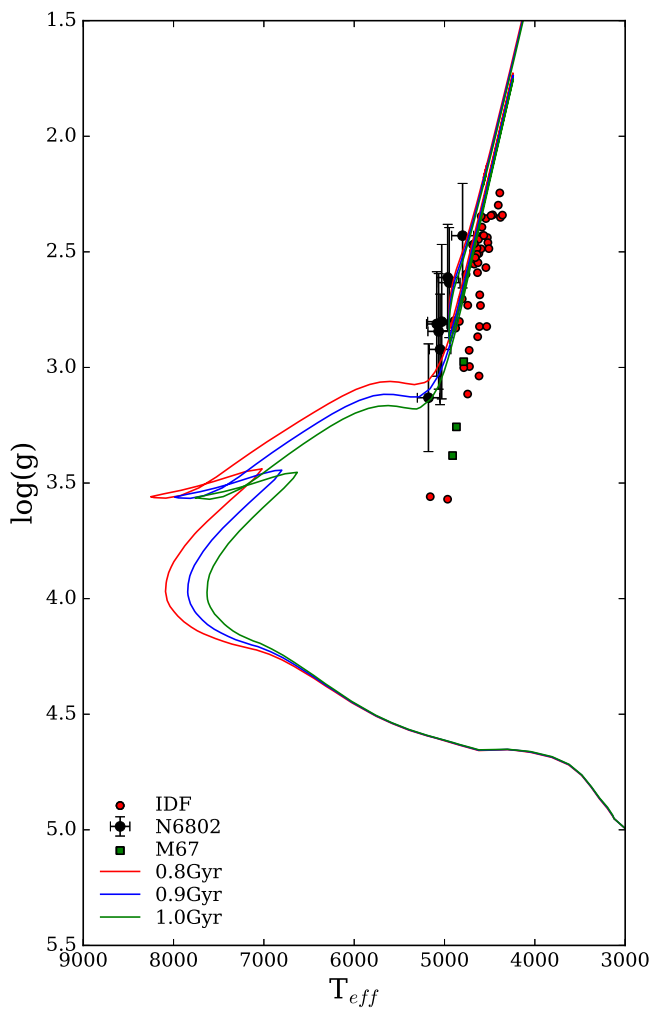} 
\caption{NGC 6802 giant cluster members with available log(g) and T$_{\rm eff}$ measurements are plotted as black dots, and the error bars indicate the associated measurement errors. Red, blue, and green solid lines represent the PARSEC isochrones with ages of 0.8, 0.9, 1.0 Gyr, respectively. For all three isochrones, [Fe/H$]=0.1$ dex. The red dots are inner disk field giants, while the green squares are M67 giant cluster members with UVES observations. See Sect.  \ref{sect:nc} for more details about the last two samples. }\label{fig:tg}
\end{figure}

Compared with the isochrone fitting in JH11, we have new [Fe/H] information in this work. We retrieved the PARSEC v1.2S isochrones \citep{Bressan2012, Chen2015} from their public website\footnote{http://stev.oapd.inaf.it/cgi-bin/cmd.}. Solar metallicity (Z$_{\odot}$) is set to be 0.0152 for this version of isochrones, so we obtained isochrones with $Z=0.0191$\footnote{We assume total metallicity is the same as iron abundance since the median [$\alpha$/Fe] of NGC 6802 stars is close solar (Sect.  \ref{sect:alpha}).}. We begin our search for age, distance, and reddening with those values from JH11. The isochrones are first shifted horizontally (by altering $E(B-V)$) or vertically (by altering $(m-M)_0$) so that the red-clump phase matches the giant cluster members. Then we pick a set of three isochrones (age step size equals 0.1 Gyr) that best covers the main-sequence turn-off stars to estimate the age and the associated errors. The middle age of the three isochrones is the estimated age, while the error is the age step. After changing the isochrones, we need to make sure the red clump phase matches the giant cluster members by repeating the procedures aforementioned. As a result, the age of NGC 6802 is estimated to be $0.9\pm0.1$ Gyr, with $(m-M)_0=11.4$ and $E(B-V)=0.86$ (Figure \ref{fig:cmd}). Our results are basically consistent with the results of JH11,  with a younger age ($-0.55$ Gyr) in this work, and $\Delta (m-M)_0=0.1$, $\Delta E(B-V)=0.02$. We note that the main-sequence turn-off cluster members are low in numbers, with a substantial photometric dispersion. This may be related to the rotation that broadens the MS turn-off \citep[e.g.,][]{DAntona2015, Li2014}, but more confirmed cluster members near the MS turn-off are needed to test this hypothesis. The PARSEC isochrone, with an age of 0.9 Gyr and a metallicity of 0.10 dex, suggests the turn-off mass (M$_{\rm TO}$) of NGC 6802 is $\sim$1.9 M$_{\odot}$.
Figure \ref{fig:tg} shows the stellar measurements of the giant cluster members and PARSEC isochrones in the log(g) versus T$_{\rm eff}$ parameter space. The dwarf member stars either have no log(g) or T$_{\rm eff}$ measurement, or have log(g) and T$_{\rm eff}$ measurements with large errors. We defer the discussion of the dwarf members to a future data release.  Though one star with the largest log(g) seems to miss the red clump, generally speaking, the stellar parameters of giant members fall closely to the red clump phase, given the uncertainties in log(g) and T$_{\rm eff}$ parameter measurements.

\section{Results}
\label{sect:re}

\subsection{Chemical pattern}
\label{sect:cp}

The individual stellar chemical abundances of NGC 6802 giants derived by GES are shown in Table \ref{tab2}. The chemical abundances of NGC 6802 dwarfs will not be discussed in this work.
The original abundances are given in the form of $\log{\epsilon(\rm {X})}$, which is defined as $\log{\rm {N_{X}/N_{H}}}+12.0$, and $\rm {X}$ is the given element. The chemical abundances are converted to a more frequently used format with the equation: $[\rm {X/Fe}]=\log{\epsilon(\rm {X})}-\log{\epsilon(\rm {X_{\odot}})}-[Fe/H]$. Because elements may have different ionized states, the [Fe/H] in this equation is in fact [\fei/H] or [\feii/H], depending on the ionized state of the element, for example, [Mg/Fe$]=[$\mgi/\fei], [Ba/Fe$]=[$\baii/\feii]. Because not all the nodes of WG11 forced ionization balance between \fei~and \feii~during the abundance derivation, this approach is more desirable. [\fei/H] is used for abundances from molecular lines.
\citealt{Grevesse2007} solar abundances are chosen here. The systematic uncertainties are minimized and accounted for in the error budget of iDR4 abundances. Therefore, using a different solar scale only produces a linear scaling effect. 

Since [X/Fe] is calculated as $\log{\epsilon(\rm {X})}-\log{\epsilon(\rm {X_{\odot}})}-\rm{[Fe/H]}$, the total error is the quadratic sum of all three contributors:
$\delta_{\rm tot}=\sqrt{\delta_{\rm{X}}^2+\delta_{\rm{X}\odot}^2+\delta_{\rm{[Fe/H]}}^2}$.
Because the  covariances\footnote{e.g., the offsets of X and Fe to the true values may have opposite signs.} between individual contributors are not accounted for, we should envision the total error as an upper limit for the measurement error, that is, it is larger than the ``true'' error.
Table \ref{tab4} lists the median abundances\footnote{ Median value is more preferable for datasets with possible outliers.}, standard deviation ($\sigma$), and upper limits for measurement errors ($\delta_{\rm tot}$). In the same table, we also show the typical number of transitions used for abundance determination for each star (Nlines). Generally speaking, elements derived from more lines are more reliable.

\begin{figure*}
\centering
\includegraphics [width=0.8\textwidth]{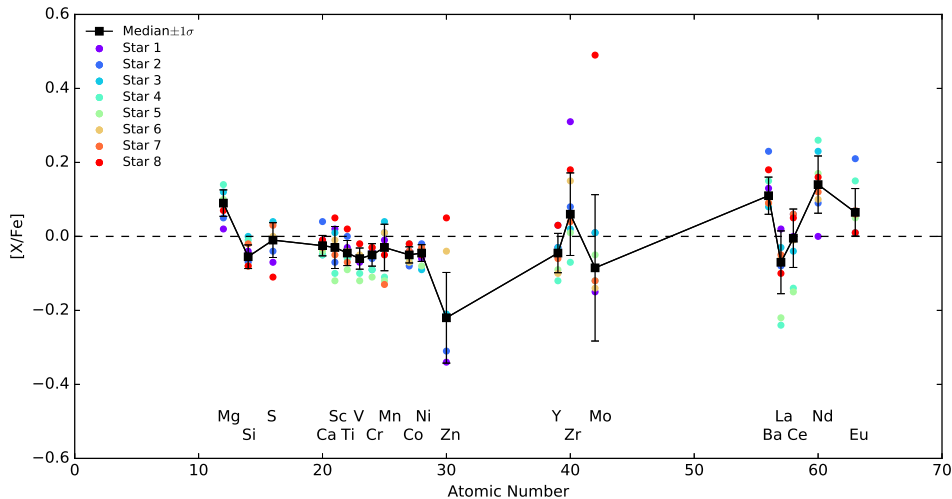} 
\caption{Element abundance as a function of atomic number for NGC 6802 giants. Cluster members are labeled with different colors, and the median abundances are shown as black squares. The error bars indicate the standard deviations of the measurements. }\label{fig:am}
\end{figure*}

Figure \ref{fig:am} shows the element abundances as a function of atomic number for NGC 6802 giants. We did not put C, N, O, Na, and Al in this figure because special treatments are needed for these elements, which will be discussed below.
The error bars indicate the median$\pm 1 \sigma$ regions. The standard deviation ($\sigma$) and upper limits for measurement errors ($\delta_{\rm tot}$) of Table \ref{tab4} suggest that most of the elements show $\sigma$ smaller than $\delta_{\rm tot}$, except Mo. Figure \ref{fig:am} indicates that the large $\sigma$ of Mo is mainly caused by one outlier (Star 8)\footnote{The Mo abundances in NGC 6802 giants are provided by only one node. An independent analysis of the spectra is needed to identify if this is a true abundance difference.}. Therefore, we conclude that there is no intrinsic scatter in NGC 6802 stars at the GES measurement level.


\begin{figure*}
\centering
\includegraphics [width=0.8\textwidth]{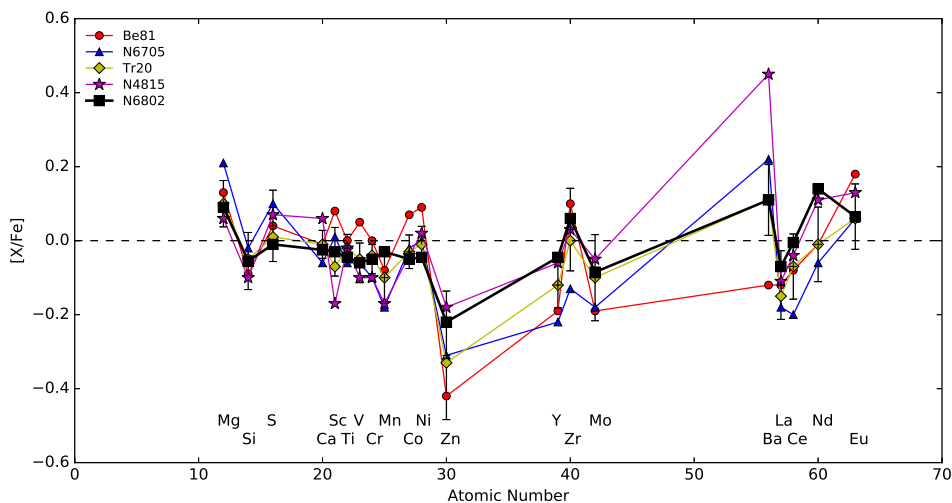} 
\caption{Median element abundance as a function of atomic number for five IOCs. Cluster names of different colors and symbols are indicated in the top left of the figure.  The error bars are median values of the abundance standard deviations of five GES IOCs.
}\label{fig:amall}
\end{figure*}


To obtain a more global perspective about IOCs, we include four previously studied GES IOCs: Be 81, NGC 6705, Tr 20, and NGC 4815. We retrieve their measurements from iDR4 and plot the median element abundances of their giant cluster members in Figure \ref{fig:amall}, analogous to Figure \ref{fig:am}. 
These four IOCs have been studied with previous data releases \citep{Magrini2015, Tautvaisiene2015, Smiljanic2016}, but iDR4 includes new elements that we will now explore for the first time, specifically, S, Mn, Co, Zn, Zr, Mo, Ba, La, Ce, and Nd.
Table \ref{tabcl} lists the basic information of these four IOCs and NGC 6802. The parameters of NGC 6802 have been discussed above. For the rest of the open clusters, E($B-V$), age, and R$_{\rm GC}$ values are given in \citet[][and reference to the original papers therein]{Magrini2015}, M$_{\rm TO}$ values are given by \citet{Smiljanic2016}, and [Fe/H] are the median metallicities of giant cluster members in iDR4. As a result, these metallicities may be slightly different from previously published values based on earlier GES data releases.

In Figure \ref{fig:amall}, we also calculate the median values of the abundance standard deviations of five GES IOCs for individual elements, which are shown as error bars around the median values of the  median abundances of five GES IOCs. These error bars in fact represent the typical abundance standard deviations for one GES IOC in individual elements. For most of the elements, the typical abundance standard deviation is similar to the cluster-to-cluster variation. An exception is found in [Ba/Fe], where the former one is significantly smaller than the latter one. Interestingly, [Ba/Fe] does seem to show unique behavior compared to other s-process elements (Sect.  \ref{sect:nc}).

\subsection{C \& N}
\label{sect:co}
C, N, and O are the most abundant metals (elements heavier than helium), and thus their study is critical for precise determination of stellar and cluster evolution. The CNO cycle is the dominant energy source in stars more massive than 1.3 M$_{\odot}$. During the first dredge-up (1DUP, \citealt{Iben1967}), C, N, Na, and some other light element abundances are modified depending on the initial stellar mass and metallicity \citep{Charbonnel1994,Boothroyd1999}. Standard stellar evolution models include only convection as a transport mechanism. However, the inclusion of other physical processes is needed to explain all abundance changes observed in red giants (see, e.g., \citealt{Pinsonneault1997}, \citealt{Smiljanic2009}, \citealt{Charbonnel2010}, \citealt{Cantiello2010}, \citealt{Tautvaisiene2013}, 2015, and references therein). Among such processes one can list, for example, rotation induced mixing, thermohaline mixing, and transport by magnetic buoyancy (see, e.g., \citealt{Busso2007}, \citealt{Charbonnel2007}, \citealt{Eggleton2008}, \citealt{Lagarde2012}).

\begin{figure}

\includegraphics [width=0.45\textwidth]{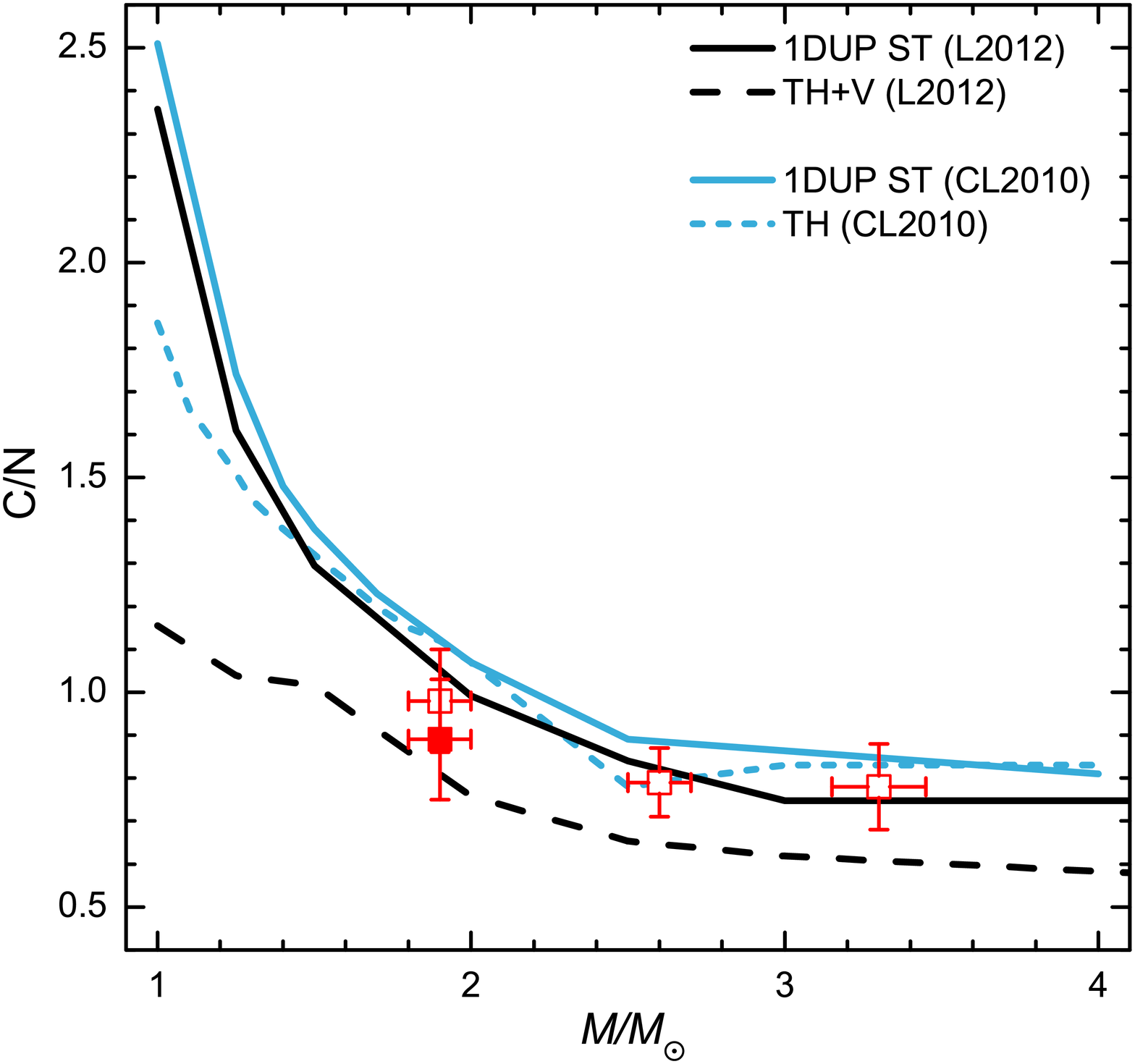} 
\caption{Mean C/N ratio of clump stars in open clusters as a function of turn-off mass. Three open clusters from T15 are labeled as open squares, while NGC 6802 is the filled square. The solid lines represent the C/N ratios predicted by the first dredge-up with standard stellar evolutionary models in \citet{Charbonnel2010} (blue solid line) or \citet{Lagarde2012} (black solid line). The blue dashed line shows the prediction when just thermohaline extra-mixing is introduced \citep{Charbonnel2010}, and the black dashed line is for the model that includes both the thermohaline and rotation induced mixing \citep{Lagarde2012}.
}\label{fig:cn}
\end{figure}

Among different mixing models \citep{Charbonnel2007, Charbonnel2010, Lagarde2012}, the C/N ratio is proposed to be a possible discriminator. By studying NGC 6705, Tr 20 and NGC 4815 from GES, \citet[][T15]{Tautvaisiene2015} concluded that the C/N ratios seem to be close to the predictions of standard models and thermohaline mixing models. Following the recipe of T15, GES derived C from C2 molecular lines, N from CN molecular lines and O from the single atomic line at 6300 \AA. 
For the giant cluster members in NGC 6802, the median [C/Fe] is $-$0.16 dex, and the median [N/Fe] is 0.51 dex, both with small standard deviation (0.05 dex). This is consistent with the depleted C and enhanced N abundances that also can be found in other IOCs (T15). 
Similar to T15, we try to distinguish different models using the C/N$-$M$_{\rm TO}$ diagram. In Figure \ref{fig:cn}, the three open clusters in T15 are shown as open squares, while NGC 6802 is shown as a filled square. 
 Measuring C/N at the red clump means that all effects from rotational mixing on the main sequence, 1DUP, rotational mixing on the RGB (red giant branch), and thermohaline mixing on the RGB in low-mass stars are included. This is a useful way to evaluate different mixing processes. 
The mean C/ N ratio of NGC 6802 lies between predictions of several models, indicating that the depleted C and enhanced N may be caused (1) by the 1DUP at the bottom of the red giant branch; (2) by the thermohaline mixing which yields indistinguishable C/N ratios as 1DUP at this M$_{\rm TO}$, or (3) by thermohaline- and rotation-induced mixing which agrees with the observed C/N  of NGC 6802 within uncertainties. Thus, 
the addition of NGC 6802 C/N ratio does not enlighten the current discussion about various models.

It is worth mentioning that NGC 6802 has been selected by the APOGEE project as one of their cluster targets \citep{Zasowski2013}. Comparing its C, N, and other element abundances from the two surveys would be useful for studying the possible systematic differences between GES and APOGEE.
 

\subsection{Na \& Al}


Na and Al are mainly synthesized by C and Ne burning in massive stars \citep{Arnett1971, Clayton2007,Woosley1995}, but they can also be modified by the NeNa, and MgAl cycles in AGB (asymptotic giant branch) stars \citep{Arnould1999,Ventura2013}.
\citet[][S16]{Smiljanic2016} analyzed a sample of 1303 stars using GES iDR2/3. Their sample consists of dwarfs and giants from both the solar neighborhood and 16 open clusters, where four previously studied IOCs are also included. S16 showed that the observed trend of [Na/Fe] (NLTE\footnote{Non-local thermodynamic equilibrium.}) and M$_{\rm TO}$ can be explained by the internal evolutionary processes, but cannot differentiate between models with and without rotation-induced mixing.

To put NGC 6802 into this picture, we first calculate the NLTE effect of our stars following the recipe of S16. The median NLTE correction for Na is $-0.13$ dex. Also we notice that the difference in $\log{\epsilon(\rm {Na_{\odot}})}$ between G07 and S16 is $ -0.13$ dex (6.17 dex in G07, and 6.30 dex in S16). After correcting for NLTE effects and changing the [Na/Fe] to the solar scale of S16, the [Na/Fe] (NLTE) in NGC 6802 is found to be 0.23$\pm$0.09 dex at M$_{\rm TO}=1.90$ M$_{\odot}$. Comparing with Figure 5 of S16, our data agree with the predictions from models with and without rotation-induced mixing, due to the small difference of the model predictions in this M$_{\rm TO}$ range. 

For Al, following S16, we assume an NLTE correction of the order of $-0.05$ dex. The difference in $\log{\epsilon(\rm {Al_{\odot}})}$ is $ -0.07$ dex (6.37 dex in G04, and 6.44 dex in S16). Stellar evolution models do not predict a change in the surface [Al/Fe] ratio in low- and intermediate-mass red giants, as the MgAl cycle is not activated in their interiors. The [Al/Fe] that we obtain for NGC 6802 is consistent with this expectation, as the median value of [Al/Fe] (NLTE) ($ 0.04 \pm 0.04$) agrees with the solar value adopted in S16 within the uncertainties.
In conclusion, the Na and Al abundances of NGC 6802 support the statement of S16, that is, that the Na and Al can be explained by the standard models, but the models with rotation cannot be excluded.

\subsection{$\alpha$ elements}
\label{sect:alpha}

According to the nucleosynthetic processes that are associated with different $\alpha$ elements during type II supernova (SNe), O and Mg are commonly classified as hydrostatic $\alpha$ elements, and Si, Ca, and Ti are classified as explosive $\alpha$ elements \citep{Woosley1995}. O and Mg are two of the primary $\alpha$ elements produced, and they are produced in almost the same ratio
for stars of disparate mass and progenitor heavy element abundance. On the other hand, two of the heaviest explosive $\alpha$ elements, Ca and Ti, follow O and Mg in the Galactic environment (e.g., Milky Way bulge), but seem to have a substantial contribution besides type II SNe in the extreme extragalactic environment \citep[e.g., massive elliptical galaxies;][]{Worthey2014,TangB2014}.

\citet{Hayden2015} analyzed a sample of 69,919 red giants from the APOGEE data release 12,  with Galactic radius $3<R_{\rm GC}<15$ kpc and height $|z|<2$ kpc. The sample of disk stars was divided into 18 bins with different Galactic radius and height. The exhaustive visualization of two sequences in the [$\alpha$/Fe] versus [Fe/H] diagram throughout 18 bins is very informative for chemical evolution studies. The stars inside 7 kpc lie in a sequence that starts from $\alpha$-rich metal-poor stars, and ends at [Fe/H]$\sim+0.4$ dex, [$\alpha$/Fe]$\sim0$ dex.

\begin{figure}

\includegraphics [width=0.5\textwidth]{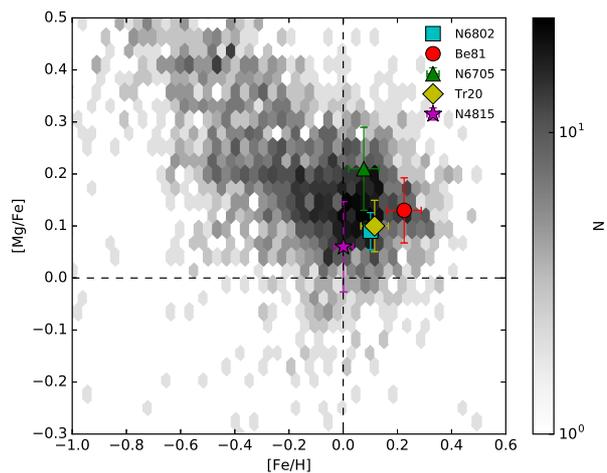} 
\caption{[Mg/Fe] vs. [Fe/H] distribution. The background color map is the number density in logarithmic scale. The text describes the selection method. Symbol meanings are indicated in the top right. }\label{fig:mg}
\end{figure}



We assemble stars in iDR4 with available Mg and Fe measurements derived from UVES spectra without further limitation, and zoom in on the region satisfying $-1.0<[$Fe/H$]<0.6$ and $-0.3<[$Fe/H$]<0.5$ in Figure \ref{fig:mg}. The number density of stars in logarithmic scale is indicated by a color map, where high number density regions have darker colors. The spectra observed by UVES have very high resolution, but the sample size is limited (2 824 stars are used in Figure \ref{fig:mg}).  We are hard-pressed to distinguish two $\alpha$-sequences with the UVES data. However, \citet{Kordopatis2015} separated two $\alpha$-sequences with the GIRAFFE abundances from GES iDR2 and additional velocity information.  The median abundances of the GES IOCs are also shown in Figure \ref{fig:mg}. The cluster abundances are quite scattered, but all fall inside the high number density region. We notice that the region with the highest number density in Figure \ref{fig:mg} is around [Mg/Fe]$\sim$0.1 dex, but the same region is around [Mg/Fe]$\sim$0 dex in other similar works \citep[e.g.][]{Hayden2015, Kordopatis2015,Wojno2016,Casey2016R}. This offset in [Mg/Fe] is likely related to the choice of the atomic data and line list for Mg. Magrini et al. (in prep.) show that using the solar Mg/Fe abundance measured by GES ($0.15$, compared to $0.08$ in G07) cancels this offset.
The readers are reminded that adopting G07 or GES solar [Mg/Fe] only produces a linear scaling effect, in other words, the relative position between the background color map and GES IOC measurements is fixed.
 

If we look at NGC 6802 cluster members closely, their $\alpha$ abundances are close to solar: [O/Fe$]=0.01\pm 0.08$, [Mg/Fe$]=0.09\pm 0.04$, [Si/Fe$]=-0.04\pm 0.03$, [Ca/Fe$]=-0.02\pm 0.03$, [Ti/Fe$]=-0.04\pm 0.04$. The mean abundance of all five $\alpha$ elements (O, Mg, Si, Ca, Ti) is $-0.02\pm0.05$ dex.   Thus, NGC 6802 is consistent with a G07 scaled-solar abundance pattern.
 
\subsection{Iron-peak elements}
NGC 6802 stars show a homogeneous iron abundance of 0.10$\pm$0.02 dex. 
The median iron abundance of NGC 6802 stars derived by GES is consistent with that of the other GES IOCs, which show super-solar iron abundances up to $+$0.23 dex. The super-solar iron abundances in IOCs agree with literature on the (negative) metallicity gradient of the Milky Way \citep[e.g.,][]{Magrini2010, Hayden2015}. Using intermediate-age or older open clusters observed by GES as tracers, \citet{Jacobson2016} suggested a radial metallicity gradient of $-0.10\pm-0.02$ dex kpc$^{-1}$ in the range $\rm R_{gc}\sim6-12$ kpc, where NGC 6802 is included in the sample.

Though Type Ia SNe, runaway deflagration obliterations of white dwarfs, have a signature more tilted towards the iron-peak group \citep{Nomoto1997}, the solar composition of the iron-peak elements are in fact a heterogeneous combination of Type Ia supernova and core collapse Type II SNe \citep{Woosley1995, Pignatari2016}.
The iron-peak elements of NGC 6802 and the other four GES IOCs show nearly solar abundances for Sc, V, Cr, Mn, Co, and Ni (Figure \ref{fig:amall}). What we find is consistent with other literature results, for example: \citet{Reddy2015} showed that their four outer disk open clusters have [V/Fe], [Cr/Fe], [Mn/Fe], [Co/Fe], and [Ni/Fe] between $\pm0.1$ dex.
However, Zn shows different characteristics:  (1) [Zn/Fe$]=-0.22\pm 0.13$ in NGC 6802. The derived [Zn/Fe] show substantial scatter\footnote{Two Zn\,{\sc i} lines at 4810 and 6362 \AA~are used in GES to derive Zn abundance. Due to the blending CN lines or the Ca\,{\sc i} lines around 6361 \AA, the Zn\,{\sc i} line at 6362 \AA~may be subject to error. Duffau et al. (submitted) compare the Zn abundances derived from only the 4810 \AA~line with those derived from both lines. The higher Zn ([Zn/Fe$]\gtrsim0$, using both lines) abundance giants disappear if the 4810 \AA~Zn abundances are used. Therefore Duffau et al. suspect that the higher [Zn/Fe] values are the result of an incorrect synthesis of the region around the Zn\,{\sc i} 6362 \AA~line in iDR4.} (Figure \ref{fig:am}); (2) similar Zn depletion is also present in the other GES IOCs. The outer disk open clusters in \citet{Reddy2015} show [Zn/Fe$]\sim-0.3$. Therefore Zn may have a  different formation mechanism, which is also the reason that recent studies separate Zn from other iron-peak elements. One of the possible mechanisms is the weak s-process (see below). 


\subsection{Neutron-capture elements}
\label{sect:nc}
Elements heavier than Fe can be synthesized via neutron-capture processes. Depending on its relative timescale of subsequent neutron capture to $\beta$-decay, the neutron-capture processes are divided into the slow (s-) and rapid (r-) processes. The main s-process elements are synthesized by AGB stars during the thermal pulsations \citep{Busso2001,Karaks2014}. Two nuclear reactions are the major neutron excess sources in AGB stars: $\rm {^{13}C(\alpha,n) ^{16}O}$ and  $\rm {^{22}Ne(\alpha,n) ^{25}Mg}$. The first reaction dominates the low mass AGB stars, while the latter one is mainly found in massive AGB stars \citep{Cristallo2015}. As the number of free neutrons per iron seed increases, for example, from $\rm {^{13}C(\alpha,n) ^{16}O}$, the s-process flow first seeds the light s-process peak (Sr$-$Y$-$Zr), extending to $\rm ^{136}Ba$, and then reaches the heavy s-process peak (Ba$-$La$-$Ce$-$Pr$-$Nd), extending to $\rm ^{204}Pb-^{207}Pb$ \citep{Bisterzo2014}. Therefore, the heavy s-process element to light s-process element ratio ([hs/ls]) is closely related to metallicity and initial stellar mass.
At even lower metallicity, where the number of free neutrons per iron seed reaches the highest value, the s-process mainly feeds the $\rm ^{208}Pb$ \citep{Bisterzo2014}. This so-called strong s-process generates 50\% of the solar Pb through low metallicity AGB stars \citep{Gallino1998}.
At the same time, the weak s-process that takes place in massive, fast evolving stars can also produce a significant amount of elements between Fe and Sr \citep{Beer1992,Pignatari2010}.
The r-process elements are generally accepted as the product of core collapse SNe \citep{Thielemann2011,Boyd2012}. Recently, however, electron capture SNe \citep{Woosley2015}, and neutron star merger \citep{Berger2013, Ji2016} are also discussed as possible sources for r-process elements.
 

\begin{figure}
\includegraphics [width=0.5\textwidth]{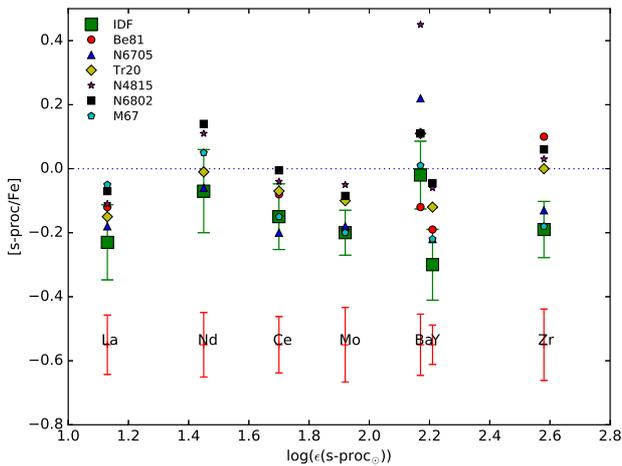} 
\caption{Cluster median [s-proc/Fe] as a function of log($\epsilon$(s-proc$_{\odot}$)), where log($\epsilon$(s-proc$_{\odot}$)) is the G07 solar abundance. The s-process element names and the median of the s-process abundance standard deviations of five GES IOCs are indicated in the bottom of the panel. The median abundances and standard deviations of the inner disk field (IDF) giants are labeled as green squares with error bars. The symbols of the five GES IOCs and M67 are indicated in the top left. The dotted line represents [s-proc/Fe$]=0$.}\label{fig:fx}
\end{figure}

\begin{figure*}
\centering

\includegraphics [width=0.8\textwidth]{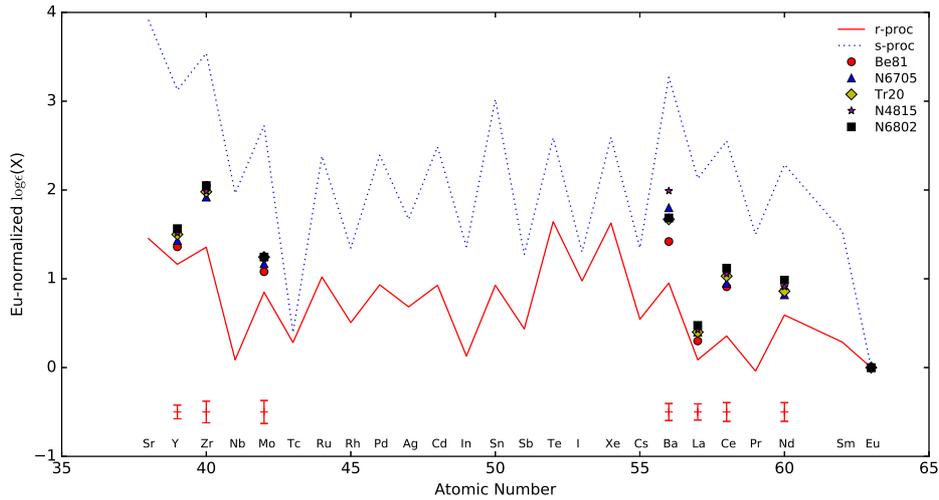} 
\caption{Eu-normalized $\log{\epsilon({\rm X})}$ as a function of atomic number for the five IOCs. The solar element r-process and s-process abundance distributions from \citet{Simmerer2004} are shown as red and blue solid lines, respectively. The element names are listed in the bottom of the panel, and the error bars indicate the standard deviations of the seven s-process elements discussed in this work.}\label{fig:ncap}
\end{figure*}

We construct internal control samples to avoid possible linear scaling effect caused by adopting different solar abundances in other studies. Firstly, the different stellar density and chemical enrichment history of the stars in clusters and in the field may show their fingerprints in s-process elements, thus we should construct a internal control sample with inner disk field (IDF) stars. From GES iDR4, we gather the IDF stars toward the Galactic bulge direction with available  UVES measurements of seven main s-process elements (Y, Zr, Mo, Ba, La, Ce, and Nd) plus Eu, and select stars within $0<{\rm [Fe/H]}<0.23$\footnote{Metallicity range for five IOCs.}. Because the stars that we study in the IOCs are mainly giants, we further posit log(g$)<3.8$ to select the giants\footnote{We do not use solar neighborhood stars as control sample, because most of these type of stars observed in GES are either dwarfs or more metal-poor than IOCs.} (see Figure \ref{fig:tg}). In total, 53 IDF giants are found in iDR4 (Figure \ref{fig:tg}), with a median metallicity of 0.09. Given that (1) the measurement errors of log(g) and T$_{\rm eff}$ are similar to the errors of NGC 6802 stars and, (2) the T$_{\rm eff}$ of isochrones with different ages change very little along the red giant branch, it is thus hard to determine the age of the IDF giants from isochrone fitting alone. However, we do see evidence that IDF giants may have different  (likely older) stellar populations than NGC 6802 and other GES IOCs. 
Secondly, recent studies are debating the existence of an anti-correlation between s-process abundance and cluster age (see Sect.  \ref{sect:dis} for more details). The star cluster M 67 is estimated to have an age of $\sim$3.5 Gyr \citep{Chen2014,Bonatto2015}, which is larger than that of the five GES IOCs (0.30$-$1.50 Gyr in Table \ref{tabcl}). A comparison of s-process abundances in IOCs and M 67 may enlighten the discussion aforementioned. There are three M 67 giants observed by GES using UVES (Figure \ref{fig:tg}). 
We analyzed the element abundances of these two control samples following the same method that we outline in Sect. \ref{sect:cp} to ensure self-consistent comparison.

In Figure \ref{fig:fx}, we plot the median abundances of seven s-process elements versus their G07 solar abundances ([s-proc/Fe] versus log($\epsilon$(s-proc$_{\odot}$)), where s-proc is one of the elements including Y, Zr, Mo, Ba, La, Ce, and Nd). The median IDF abundances are labeled as green squares, while the associated error bars represent the standard deviations. On the same figure, we also plot median s-process abundances of different open clusters, including M 67. Figure \ref{fig:fx} indicates that the main s-process abundances in IOCs show more cluster-to-cluster variation than other elements, for example, the [Ba/Fe] difference between NGC 4815 and Be 81 is $\sim$0.6 dex (also see Figure \ref{fig:amall}).  We also see that the IOC abundances seem to be systematically higher than that of the IDF. Particularly, all seven s-process abundances of NGC 6802 are more than 1$\sigma$ larger than the IDF s-process abundances. Meanwhile, M67 stars also seem to have systematically lower Y, Zr, Mo, Ba, and Ce abundances than that of the IOCs. We will further explore the implications in Sect.  \ref{sect:dis}.

To quantify the efficiency of main s-process in IOCs, we compare the seven s-process elements in IOCs with solar elemental r-process and s-process abundance distributions published by \citet{Simmerer2004} (Figure \ref{fig:ncap}). The abundances are given in spectroscopic units, $\log \epsilon(\rm{X})$, which is the same as that used in GES. 
We normalize the abundance distributions by Eu, since it is primarily an r-process element \citep[97\% r-process,][]{Simmerer2004}. We find that all the s-process elements fall between the solar s-process and r-process distributions. However, the seven s-process abundances have different relative distances to the solar r-process (or s-process) distribution. 
This indicates at face value that the s-process elements in IOCs are generated by a different percentage combination of the s-process and r-process compared to that of the Sun, even when standard deviations are considered (the error bars at the bottom of Figure \ref{fig:ncap}), but it also can be caused by the uncertainties associated with the solar r-process and s-process distributions. We note that this experiment has nothing to do with identifying the s-process or r-process origin of the elements.

\begin{figure}
\includegraphics [width=0.5\textwidth]{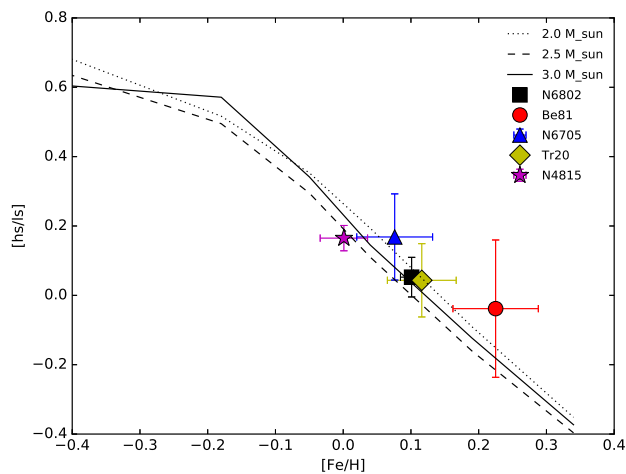} 
\caption{Heavy to light s-process element ratio ([hs/ls]) as a function of metallicity. Typical standard deviations are indicated by the error bars. The three AGB models from \citet{Cristallo2015} are shown with different line styles.}\label{fig:hsls}
\end{figure}

According to the nucleosynthetic mechanisms mentioned above, the heavy to light s-process element ratio ([hs/ls]) is an indicator of metallicity and initial stellar mass. We define [ls/Fe] as ([Y/Fe]$+$[Zr/Fe])/2, [hs/Fe] as ([Ba/Fe]$+$[La/Fe]$+$[Nd/Fe])/3, and [hs/ls] is equal to [hs/Fe]$-$[ls/Fe]. 
To guide our understanding, [hs/ls] are compared with that of the AGB models from \citet{Cristallo2015} as a function of metallicity  (Figure \ref{fig:hsls}). AGB models with 2, 2.5, and 3 M$_{\odot}$ stellar mass are chosen to illustrate the general trend. \citet{Lodders2003} solar abundances are used in these AGB models, so we shift the solar abundances of these models to that of G07. But there is still a linear scaling offset between our results and the models, so we arbitrarily shift $+$0.1 dex for [Fe/H] in the models.  Since metallicity is generally used as a clock, one possible explanation for the abundance shift is that the AGB model metallicity ought to be lower than the metallicity of the stars that formed from material enriched by its winds. This shift does not affect our conclusions below, because we are only interested in the general trend of [hs/ls] as a function of [Fe/H], instead of the absolute values. 
We see that different initial mass AGB models only show small separation in the metal rich region.
The anti-correlation between [hs/ls] and metallicity are found in IOCs and in AGB models. Therefore our results in fact support the nucleosynthetic notion that [hs/ls] is a good indicator of metallicity in the metal-rich region.
The numbers in Figure \ref{fig:hsls} should not be applied blindly by the users who are interested in absolute values.

\section{S-process elements in IOCs, M 67, and IDF}
\label{sect:dis}

The element abundances [Ba/Fe] in young open clusters are found to be enhanced \citep[e.g.,][]{DOrazi2009,Yong2012,Jacobson2013}, where an anti-correlation between [Ba/Fe] and cluster age is also suggested. Except Be 81, four GES IOCs have [Ba/Fe] higher than that of M 67, which agrees with the anti-correlation, since M 67 is older than GES IOCs.  The situation concerning other s-process elements is less certain in the literature. \citet[][M11]{Maiorca2011} reported the discovery of enhanced s-process element abundances, including Y, Zr, La, and Ce, in open clusters younger than the Sun. However, there are also contradicting results in the literature. For example, \citet{Jacobson2013} showed no clear enhancement in [La/Fe] and [Zr/Fe] in a sample of 19 open clusters,  and \citet{Yong2012} also found no clear enhancement in [Zr/Fe] in a sample of five open clusters. Given that our five IOCs have ages between 0.30 and 1.50 Gyr (Table \ref{tabcl}), the enhanced [Y/Fe], [Zr/Fe], [Mo/Fe], and [Ce/Fe] in IOCs compared to that of M 67 stars in this work (Sect.  \ref{sect:nc}) may support the discovery of M11. However, we note that the opposite is found in [La/Fe]. Because the cluster age range of our IOCs is small compared to that in the literature works, our data alone is not as suitable for further discussing the relation between element abundance and cluster age. However, it will be interesting to verify this relation when more open clusters with a wider range of (self-consistent) ages are available in GES.

Next, we investigate the s-process abundances in IOCs, M67, and IDF together. Similar to M 67, four GES IOCs (except Be81) have [Ba/Fe] more than 1$\sigma$ higher than that of the IDF stars. As for the other six s-process elements, most of the IOC (median) [s-proc/Fe] are higher than the IDF (median) [s-proc/Fe]. However, only NGC 6802 and NGC 4815 show 1$\sigma$ higher [s-proc/Fe] than that of IDF stars in all seven s-process elements. 
On the other hand, Y, Zr, Mo, Ba, Ce, and Nd abundances of M67 and IDF agree within the IDF uncertainties. Given that Figure \ref{fig:tg} indicates different stellar populations between IDF and IOCs, while the stellar populations of M 67 and IDF seem more consistent, our results should be interpreted as the followings: (1) comparing IOCs and IDF may not lead to conclusions about cluster and field environments as they may have different stellar populations; (2) supposing that IDF stars have similar ages to M67, the closely matched [s-proc/Fe] in M 67 and in IDF suggest that the s-process difference between the cluster and field environments may be very small. This nicely fits into the picture that field stars are dissolved from open clusters \citep[e.g.,][]{Freeman2002,Pavani2007}; (3) the result that IOC median [s-proc/Fe] are larger than that of M 67 agrees with the hypothesis that [s-proc/Fe] anti-correlates with cluster age.
 
While studies of s-process elements from more GES data or other independent data are still required to firmly establish the anti-correlation between [s-proc/Fe] and cluster age, theoretical explanations may enlighten the discussion from a different direction. For example,
\citet{Maiorca2012} suggested the anti-correlation between s-process abundance and cluster age may be explained if the number of neutrons from the $\rm {^{13}C(\alpha,n) ^{16}O}$ reaction in reservoirs are larger than previously assumed in low mass AGB stars (${\rm M} <1.5$ M$_{\odot}$). This gives a physical support for the anti-correlation. A careful nucleosynthesis study is beyond the scope of this paper and readers are referred to the literature that we cite for more discussions.

\section{Summary}
\label{sect:con}

We analyzed the kinematic properties and chemical pattern of NGC 6802 in this work, where the latter has no previous research. Using the radial velocities derived by GES, we identified eight giant cluster members from UVES measurements with a median RV of 12.0 km s$^{-1}$; and 95 dwarf cluster members from GIRAFFE measurements with a median RV of 13.6 km s$^{-1}$. Given that the medium metallicity derived from UVES observed giants is $0.10\pm0.02$ dex, the age of NGC 6208 is estimated to be $0.9\pm0.1$ Gyr with $(m-M)_0=11.4$ and $E(B-V)=0.86$.

To gain a more general picture, we supplemented NGC 6802 chemical abundances with those of four other GES IOCs (Be 81, NGC 6705, Tr 20, and NGC 4815) in a self-consistent way. Twenty-seven element species were derived in GES iDR4 from the high resolution UVES spectra; in particular, S, Mn, Co, Zn, Zr, Mo, Ba, La, Ce, and Nd are explored for the first time. Most of the elements lighter than Ni show abundances within $\pm0.1$ dex of the G07 solar abundances, while heavier elements show substantial deviation from the G07 solar values. 

Both C and N, as well as the NLTE-corrected Na and Al abundances agree with other IOCs that have been studied in T15 and S16.
Among the five $\alpha$ elments, that is, O, Mg, Si, Ca, and Ti,
NGC 6802 stars show a mean $\alpha$ abundance very close to solar. We constructed one internal control sample of 53 inner disk field (IDF) giants towards Galactic bulge, and another one that consists of three giants from M 67. Compared with M 67, GES IOCs seem to have higher [Ba/Fe], and less-certainly, higher abundances in Y, Zr, Mo, La, Ce, and Nd, which agrees with the hypothesis that [s-proc/Fe] anti-correlates with cluster age. Furthermore, we see [s-proc/Fe] in M 67 and IDF match closely, indicating that the environmental influence in s-process may be weak.


\section{Acknowledgements} 

Based on data products from observations made with ESO Telescopes at the La Silla Paranal Observatory under programme ID 188.B-3002. These data products have been processed by the Cambridge Astronomy Survey Unit (CASU) at the Institute of Astronomy, University of Cambridge, and by the FLAMES/UVES reduction team at INAF/Osservatorio Astrofisico di Arcetri. These data have been obtained from the Gaia-ESO Survey Data Archive, prepared and hosted by the Wide Field Astronomy Unit, Institute for Astronomy, University of Edinburgh, which is funded by the UK Science and Technology Facilities Council.

This work was partly supported by the European Union FP7 programme through ERC grant number 320360 and by the Leverhulme Trust through grant RPG-2012-541. We acknowledge the support from INAF and Ministero dell' Istruzione, dell' Universit\`a' e della Ricerca (MIUR) in the form of the grant "Premiale VLT 2012". The results presented here benefit from discussions held during the Gaia-ESO workshops and conferences supported by the ESF (European Science Foundation) through the GREAT Research Network Programme.

We thank the anonymous referee for insightful comments. D.G. , S.V., and B.T. gratefully acknowledges support from the Chilean BASAL   Centro de Excelencia en Astrof\'{i}sica y Tecnolog\'{i}as Afines (CATA) grant PFB-06/2007.
C.M. acknowledges support
from CONICYT-PCHA/Doctorado Nacional/2014-21141057.
R.E.C. acknowledges funding from Gemini-CONICYT for Project 32140007.
F.M. gratefully acknowledges the support provided by Fondecyt for project
314017.
This work was partly supported (A.R.C.) by the European Union FP7 programme through grant number 320360.
This work was partly supported (A.D., G.T., R.\v{Z}.) by the grant from
the Research Council of Lithuania (MIP-082/2015).
S.G.S. acknowledge the support from FCT through Investigador FCT contract of reference IF/00028/2014 and the support from FCT through the project PTDC/FIS-AST/7073/2014.
C.L. gratefully acknowledges financial support from the European Research Council (ERC-CoG-646928, Multi-Pop, PI: N. Bastian).
V.A. acknowledges the support from the FCT (Portugal) in the form of the grant SFRH/BPD/70574/2010, the support by FCT through national funds (ref. PTDC/FIS-AST/7073/2014 and ref. PTDC/FIS-AST/1526/2014) and by FEDER through COMPETE2020 (ref. POCI-01-0145-FEDER-016880 and ref. POCI-01-0145-FEDER-016886).

\bibliographystyle{aa}
\bibliography{ref2015}
\clearpage

\begin{table*}
\caption{NGC 6802 giant cluster members observed by UVES.}              
\label{tab1}      
\centering                                      
\begin{tabular}{c c c c c c c c c c}          
\hline\hline                        
\# & Cname & $V$  & $V-I$ & T$_{\rm {eff}}$  & $\log(\rm g)$  &$\xi$ & [Fe/H] & V$_{\rm rad}$  & S/N\\    
 & & (mag) & (mag) & (K) & (dex) & (km s$^{-1}$)  & (dex)  & (km s$^{-1}$)  &  \\
\hline                                   
      1 & 19303058+2013163 & 14.960 & 2.056 & 5086 $\pm$ 114 &  2.81 $\pm$  0.23 &  1.55 $\pm$  0.22  &  0.11 $\pm$  0.10  & 11.56 $\pm$  0.57 & 83 \\
2 & 19303085+2016274 & 14.716 & 2.043 & 5031 $\pm$ 156 &  2.80 $\pm$  0.33 &  1.59 $\pm$  0.05  &  0.08 $\pm$  0.12  & 12.29 $\pm$  0.57 & 108 \\
3 & 19303184+2014459 & 14.884 & 2.037 & 5049 $\pm$ 119 &  2.92 $\pm$  0.24 &  1.61 $\pm$  0.19  &  0.13 $\pm$  0.10  & 12.03 $\pm$  0.57 & 109 \\
4 & 19303309+2015442 & 14.360 & 2.095 & 4800 $\pm$ 121 &  2.43 $\pm$  0.23 &  1.66 $\pm$  0.07  &  0.08 $\pm$  0.10  & 11.46 $\pm$  0.57 & 131 \\
5 & 19303611+2016329 & 14.667 & 1.980 & 4965 $\pm$ 121 &  2.61 $\pm$  0.23 &  1.59 $\pm$  0.08  &  0.10 $\pm$  0.10  & 12.49 $\pm$  0.57 & 119 \\
6 & 19303884+2014005 & 14.985 & 1.997 & 5065 $\pm$ 119 &  2.84 $\pm$  0.25 &  1.54 $\pm$  0.11  &  0.12 $\pm$  0.10  & 13.52 $\pm$  0.57 & 107 \\
7 & 19303943+2015237 & 14.535 & 2.004 & 4948 $\pm$ 120 &  2.63 $\pm$  0.24 &  1.64 $\pm$  0.11  &  0.10 $\pm$  0.11  & 10.40 $\pm$  0.57 & 114 \\
8 & 19304170+2015224 & 14.920 & 1.990 & 5178 $\pm$ 123 &  3.13 $\pm$  0.23 &  1.49 $\pm$  0.04  &  0.09 $\pm$  0.10  & 11.54 $\pm$  0.57 & 106 \\
\hline                                             
\end{tabular}
\tablefoot{GES object name from coordinates (Cname) is formed from the coordinates of the object splicing the RA in hours, minutes, and seconds (to 2 decimal places) and the Dec in degrees,  arcminutes, and arcseconds (to 1 decimal place) together, including a sign for the declination.}
\end{table*}

\begin{table*}
\caption{Non-members observed by UVES.}              
\label{tabnon}      
\centering                                      
\begin{tabular}{c c c c c c c c c c}          
\hline\hline                        
\# & Cname & $V$  & $V-I$ & T$_{\rm {eff}}$  & $\log(\rm g)$  &$\xi$ & [Fe/H] & V$_{\rm rad}$  & S/N\\    
 & & (mag) & (mag) & (K) & (dex) & (km s$^{-1}$)  & (dex)  & (km s$^{-1}$)  &  \\
\hline 
N1 & 19303274+2014498 & 14.724 & 2.050 & 4941 $\pm$ 119 &  2.62 $\pm$  0.23 &  1.60 $\pm$  0.16  &  0.03 $\pm$  0.10  & 11.88 $\pm$  0.10 &   ... \\
N2 & 19303773+2016196 & 15.044 & 2.065 & 4984 $\pm$ 117 &  2.88 $\pm$  0.22 &  1.48 $\pm$  0.07  &  0.00 $\pm$  0.09  & 16.62 $\pm$  0.57 &    87 \\
N3 & 19303970+2013474 & 14.144 & 2.041 & 5143 $\pm$ 127 &  2.33 $\pm$  0.24 &  1.89 $\pm$  0.08  &  0.04 $\pm$  0.10  & 24.05 $\pm$  0.57 &   153 \\
N4 & 19304281+2016107 & 15.394 & 2.074 & 4766 $\pm$ 121 &  2.63 $\pm$  0.25 &  1.33 $\pm$  0.18  & -0.10 $\pm$  0.10  & 17.35 $\pm$  0.57 &    67 \\
N5 & 19304646+2015140 & 15.030 & 2.009 & 4709 $\pm$ 122 &  2.90 $\pm$  0.24 &  1.40 $\pm$  0.19  &  0.00 $\pm$  0.10  & 61.81 $\pm$  0.57 &    92 \\
\hline                                             
\end{tabular}
\tablefoot{The meaning of Cname is explained in Table \ref{tab1}.}
\end{table*}

\clearpage

\begin{table}
\caption{NGC 6802 dwarf cluster members in the lower main sequence.}              
\label{tablms}      
\setlength{\tabcolsep}{3pt} 
\centering                                      
\begin{tabular}{l c c c  r }          
\hline\hline                        
\# & Cname & $V$  & $V-I$ & V$_{\rm rad}$  \\    
 & & (mag) & (mag) & (km s$^{-1}$) \\
\hline                                   
1 & 19302019+2016175 & 18.553 & 1.736 & 14.91 $\pm$  1.94 \\
2 & 19302069+2017546 & 17.045 & 1.464 & 16.36 $\pm$  0.62 \\
3 & 19302141+2014181 & 17.307 & 1.416 & 13.73 $\pm$  1.17 \\
4 & 19302237+2013203 & 18.623 & 1.679 & 13.64 $\pm$  0.75 \\
5 & 19302283+2013079 & 17.154 & 1.526 & 11.32 $\pm$  0.77 \\
.. & ... & ... & ... & ... \\
\hline                                             
\end{tabular}
\tablefoot{Full table online.}
\end{table}

\begin{table}
\caption{NGC 6802 dwarf cluster members in the upper main sequence.}              
\label{tabums}      
\setlength{\tabcolsep}{3pt} 
\centering                                      
\begin{tabular}{l c c c r}          
\hline\hline                        
\# & Cname & $V$ & $V-I$ & V$_{\rm rad}$  \\    
   &       & (mag)& (mag) & (km s$^{-1}$)   \\
\hline                                   
1 & 19302315+2013406 & 15.386 & 1.237  & 16.23 $\pm$  1.13   \\
2 & 19302489+2016207 & 16.056 & 1.332  & 16.13 $\pm$  2.84   \\
3 & 19302535+2015250 & 17.677 & 1.491  & 14.78 $\pm$  1.23   \\
4 & 19302605+2011448 & 16.718 & 1.424  &  9.85 $\pm$  5.01   \\
5 & 19302638+2017520 & 17.228 & 1.481  & 11.24 $\pm$  0.58   \\
.. & ... & ... & ... & ... \\
\hline                                             
\end{tabular}
\tablefoot{Full table online.}
\end{table}

\begin{table}
\caption{Non-members in the lower main sequence.}              
\label{tab:lmsno}      
\setlength{\tabcolsep}{3pt} 
\centering                                      
\begin{tabular}{l c c c r}          
\hline\hline                        
\# & Cname & $V$ & $V-I$ & V$_{\rm rad}$  \\    
   &       & (mag)& (mag) & (km s$^{-1}$)   \\
\hline                                   
1 & 19301867+2014348 & 17.044 & 1.504 &  4.01 $\pm$  1.17 \\
2 & 19301894+2016438 & 17.731 & 1.486 &  9.22 $\pm$  0.89 \\
3 & 19301932+2015503 & 17.703 & 1.553 & 21.27 $\pm$  0.74 \\
4 & 19302001+2013368 & 17.592 & 1.628 & -1.83 $\pm$  0.36 \\
5 & 19302347+2018574 & 17.343 & 1.462 & 27.98 $\pm$  0.68 \\
.. & ... & ... & ... & ... \\
\hline                                             
\end{tabular}
\tablefoot{Full table online.}
\end{table}

\begin{table}
\caption{Non-members in the upper main sequence.}              
\label{tab:umsno}      
\setlength{\tabcolsep}{3pt} 
\centering                                      
\begin{tabular}{r c c c r c}          
\hline\hline                        
\# & Cname & $V$ & $V-I$ & V$_{\rm rad}$  & sp?$^{a}$\\    
   &       & (mag)& (mag) & (km s$^{-1}$)   \\
\hline                                   
1 & 19301861+2015280 & 16.239 & 1.241  & -1.38 $\pm$  0.82 &    \\
2 & 19301909+2013571 & 16.992 & 1.510  & 21.59 $\pm$  0.20 &    \\
3 & 19301985+2013424 & 15.464 & 1.193  & 18.40 $\pm$  1.70 & Y  \\
4 & 19302064+2016238 & 16.759 & 1.435  & 23.33 $\pm$  6.68 &    \\
5 & 19302120+2018430 & 17.144 & 1.390  & 61.96 $\pm$ 20.98 &    \\
.. & ... & ... & ... & ... & .. \\
\hline                                             
\end{tabular}
\tablefoot{
$^a$ Stars belong to the second peak of the radial velocity distribution are labeled with `Y'. 
Full table online.]
}
\end{table}

\begin{table*}
\caption{iDR4 recommended abundances for NGC 6802 giant cluster members.}              
\label{tab2}      
\setlength{\tabcolsep}{3pt}
\centering                                      
\begin{tabular}{c c c c c c c c c c}          
\hline\hline                        
  \#      &     C\_C2 & N\_CN   &  O\,{\sc i}  &  Na\,{\sc i}  &  Mg\,{\sc i}  &  Al\,{\sc i} &   Si\,{\sc i}   &  S\,{\sc i}  &  Ca\,{\sc i}    \\

\hline
 1  & 8.34  $\pm$ 0.01  & 8.36  $\pm$ 0.09  & 8.77  $\pm$ 0.05  & 6.63  $\pm$ 0.06  & 7.61  $\pm$ 0.12  & 6.63  $\pm$ 0.07  & 7.53  $\pm$ 0.07  & 7.13  $\pm$ 0.07  & 6.36  $\pm$ 0.08 \\
 2  &  ...  & ...  &  ...  & 6.75  $\pm$ 0.06  & 7.67  $\pm$ 0.12  & 6.64  $\pm$ 0.07  & 7.53  $\pm$ 0.07  & 7.19  $\pm$ 0.07  & 6.44  $\pm$ 0.08 \\
 3  & 8.29  $\pm$ 0.04  & 8.44  $\pm$ 0.08  & 8.90  $\pm$ 0.05  & 6.69  $\pm$ 0.06  & 7.72  $\pm$ 0.12  & 6.57  $\pm$ 0.07  & 7.58  $\pm$ 0.07  & 7.25  $\pm$ 0.07  & 6.33  $\pm$ 0.09 \\
 4  & 8.23  $\pm$ 0.01  & 8.34  $\pm$ 0.07  & 8.68  $\pm$ 0.05  & 6.83  $\pm$ 0.06  & 7.72  $\pm$ 0.12  & 6.61  $\pm$ 0.07  & 7.55  $\pm$ 0.07  & 7.18  $\pm$ 0.07  & 6.31  $\pm$ 0.08 \\
 5  & 8.28  $\pm$ 0.04  & 8.33  $\pm$ 0.08  & 8.68  $\pm$ 0.10  & 6.78  $\pm$ 0.06  & 7.73  $\pm$ 0.12  & 6.64  $\pm$ 0.07  & 7.53  $\pm$ 0.07  & 7.23  $\pm$ 0.07  & 6.37  $\pm$ 0.09 \\
 6  & 8.36  $\pm$ 0.01  & 8.29  $\pm$ 0.01  & 8.71  $\pm$ 0.10  & 6.65  $\pm$ 0.05  & 7.66  $\pm$ 0.12  & 6.59  $\pm$ 0.07  & 7.47  $\pm$ 0.07  & 7.18  $\pm$ 0.07  & 6.33  $\pm$ 0.08 \\
 7  & 8.31  $\pm$ 0.01  & 8.35  $\pm$ 0.05  & 8.70  $\pm$ 0.05  & 6.75  $\pm$ 0.06  & 7.68  $\pm$ 0.12  & 6.64  $\pm$ 0.07  & 7.55  $\pm$ 0.07  & 7.23  $\pm$ 0.07  & 6.34  $\pm$ 0.08 \\
 8  & 8.32  $\pm$ 0.01  & 8.41  $\pm$ 0.10  & 8.81  $\pm$ 0.10  & 6.58  $\pm$ 0.08  & 7.69  $\pm$ 0.12  & 6.61  $\pm$ 0.07  & 7.52  $\pm$ 0.07  & 7.12  $\pm$ 0.07  & 6.39  $\pm$ 0.08 \\
\hline                                             
  \#    &  Sc\,{\sc i}   &   Ti\,{\sc i}   &  V\,{\sc i}  &  Cr\,{\sc i}  &  Mn\,{\sc i} &  Fe\,{\sc i}  &  Fe\,{\sc ii}   &  Co\,{\sc i}  &  Ni\,{\sc i}      \\
\hline
 1  & 3.25  $\pm$ 0.07  & 4.93  $\pm$ 0.08  & 3.99  $\pm$ 0.09  & 5.66  $\pm$ 0.12  & 5.44  $\pm$ 0.17  & 7.51  $\pm$ 0.09  & 7.52  $\pm$ 0.08  & 4.94  $\pm$ 0.10  & 6.23  $\pm$ 0.11 \\
 2  & 3.19  $\pm$ 0.10  & 4.99  $\pm$ 0.08  & 4.03  $\pm$ 0.09  & 5.67  $\pm$ 0.12  & 5.49  $\pm$ 0.13  & 7.54  $\pm$ 0.09  & 7.52  $\pm$ 0.09  & 4.93  $\pm$ 0.10  & 6.30  $\pm$ 0.11 \\
 3  & 3.25  $\pm$ 0.07  & 4.92  $\pm$ 0.08  & 4.01  $\pm$ 0.09  & 5.62  $\pm$ 0.12  & 5.50  $\pm$ 0.14  & 7.52  $\pm$ 0.09  & 7.60  $\pm$ 0.08  & 4.92  $\pm$ 0.10  & 6.21  $\pm$ 0.10 \\
 4  & 3.12  $\pm$ 0.10  & 4.89  $\pm$ 0.08  & 3.95  $\pm$ 0.09  & 5.60  $\pm$ 0.11  & 5.33  $\pm$ 0.13  & 7.50  $\pm$ 0.10  & 7.47  $\pm$ 0.09  & 4.95  $\pm$ 0.10  & 6.24  $\pm$ 0.09 \\
 5  & 3.15  $\pm$ 0.10  & 4.91  $\pm$ 0.08  & 3.98  $\pm$ 0.09  & 5.63  $\pm$ 0.12  & 5.37  $\pm$ 0.12  & 7.55  $\pm$ 0.09  & 7.50  $\pm$ 0.09  & 4.96  $\pm$ 0.09  & 6.25  $\pm$ 0.10 \\
 6  & 3.20  $\pm$ 0.10  & 4.90  $\pm$ 0.08  & 3.99  $\pm$ 0.09  & 5.65  $\pm$ 0.11  & 5.44  $\pm$ 0.17  & 7.49  $\pm$ 0.09  & 7.52  $\pm$ 0.09  & 4.89  $\pm$ 0.10  & 6.23  $\pm$ 0.11 \\
 7  & 3.18  $\pm$ 0.10  & 4.89  $\pm$ 0.08  & 4.00  $\pm$ 0.09  & 5.67  $\pm$ 0.12  & 5.32  $\pm$ 0.14  & 7.51  $\pm$ 0.09  & 7.52  $\pm$ 0.08  & 4.94  $\pm$ 0.10  & 6.26  $\pm$ 0.11 \\
 8  & 3.31  $\pm$ 0.08  & 5.01  $\pm$ 0.08  & 4.07  $\pm$ 0.10  & 5.70  $\pm$ 0.12  & 5.43  $\pm$ 0.13  & 7.54  $\pm$ 0.09  & 7.60  $\pm$ 0.08  & 4.99  $\pm$ 0.10  & 6.27  $\pm$ 0.11 \\
\hline
  \#    &   Zn\,{\sc i}    &   Y\,{\sc ii}  &  Zr\,{\sc i}  &  Mo\,{\sc i}  &  Ba\,{\sc ii}  &  La\,{\sc ii} &   Ce\,{\sc ii}  & Nd\,{\sc ii} &   Eu\,{\sc ii}\\
\hline
 1  & 4.32  $\pm$ 0.13  & 2.25  $\pm$ 0.12  & 2.95  $\pm$ 0.13  & 1.83  $\pm$ 0.10  & 2.37  $\pm$ 0.14  & 1.22  $\pm$ 0.15  & 1.77  $\pm$ 0.14  & 1.52  $\pm$ 0.16  & 0.66  $\pm$ 0.17 \\
 2  & 4.38  $\pm$ 0.13  & 2.31  $\pm$ 0.12  & 2.75  $\pm$ 0.14  & 2.02  $\pm$ 0.10  & 2.47  $\pm$ 0.14  & 1.12  $\pm$ 0.15  & 1.83  $\pm$ 0.14  & 1.61  $\pm$ 0.18  & 0.80  $\pm$ 0.12 \\
 3  & 4.46  $\pm$ 0.13  & 2.33  $\pm$ 0.10  & 2.67  $\pm$ 0.13  & 2.00  $\pm$ 0.10  & 2.40  $\pm$ 0.14  & 1.25  $\pm$ 0.15  & 1.81  $\pm$ 0.15  & 1.83  $\pm$ 0.16  & 0.68  $\pm$ 0.17 \\
 4  & 4.43  $\pm$ 0.13  & 2.11  $\pm$ 0.11  & 2.56  $\pm$ 0.15  & 1.85  $\pm$ 0.10  & 2.34  $\pm$ 0.14  & 0.91  $\pm$ 0.15  & 1.58  $\pm$ 0.13  & 1.73  $\pm$ 0.15  & 0.69  $\pm$ 0.10 \\
 5  & 4.48  $\pm$ 0.13  & 2.17  $\pm$ 0.10  & 2.69  $\pm$ 0.13  & 1.97  $\pm$ 0.10  & 2.31  $\pm$ 0.14  & 0.96  $\pm$ 0.15  & 1.60  $\pm$ 0.13  & 1.67  $\pm$ 0.15  & 0.62  $\pm$ 0.17 \\
 6  & 4.60  $\pm$ 0.13  & 2.18  $\pm$ 0.11  & 2.77  $\pm$ 0.14  & 1.82  $\pm$ 0.10  & 2.33  $\pm$ 0.14  & 1.14  $\pm$ 0.15  & 1.76  $\pm$ 0.14  & 1.62  $\pm$ 0.17  & 0.66  $\pm$ 0.16 \\
 7  & 4.44  $\pm$ 0.13  & 2.22  $\pm$ 0.11  & 2.68  $\pm$ 0.15  & 1.86  $\pm$ 0.10  & 2.33  $\pm$ 0.14  & 1.15  $\pm$ 0.15  & 1.83  $\pm$ 0.15  & 1.64  $\pm$ 0.17  & 0.65  $\pm$ 0.16 \\
 8  & 4.74  $\pm$ 0.12  & 2.39  $\pm$ 0.13  & 2.85  $\pm$ 0.13  & 2.50  $\pm$ 0.58  & 2.50  $\pm$ 0.14  & 1.18  $\pm$ 0.15  & 1.90  $\pm$ 0.16  & 1.76  $\pm$ 0.15  & 0.68  $\pm$ 0.16 \\
\hline
\end{tabular}
\tablefoot{Cname can be found in Table \ref{tab1}. Abundances are given in the form of $\log{\epsilon(\rm {X})}$, which is defined as $\log{\rm {N_{X}/N_{H}}}+12.0$, and $\rm {X}$ is the given element.}
\end{table*}

\begin{table*}
\caption{Median abundance, standard deviation, total error and number of lines used for NGC 6802 giant cluster members.}              
\label{tab4}      
\centering                                      
\begin{tabular}{l r c c r}          
\hline\hline  
   &Median& $\sigma^a$&$\delta_{\rm {tot}}$& Nlines$^b$\\
\hline
 \rm [Fe\,{\sc i}/H]   &  0.07  &  0.02  & ---  &169 \\
 \rm [Fe\,{\sc ii}/H]   &  0.07  &  0.05  & ---  & 28 \\
 $\rm           [C/Fe] $  & -0.16  &  0.05  &  0.10  &  2 \\
 $\rm           [N/Fe] $  &  0.51  &  0.05  &  0.13  &  3 \\
 $\rm           [O/Fe] $  &  0.01  &  0.08  &  0.11  &  1 \\
 $\rm          [Na/Fe] $  &  0.49  &  0.09  &  0.12  &  5 \\
 $\rm          [Mg/Fe] $  &  0.09  &  0.04  &  0.17  &  5 \\
 $\rm          [Al/Fe] $  &  0.18  &  0.03  &  0.13  &  3 \\
 $\rm          [Si/Fe] $  & -0.04  &  0.03  &  0.12  &  8 \\
 $\rm           [S/Fe] $  & -0.01  &  0.05  &  0.12  &  1 \\
 $\rm          [Ca/Fe] $  & -0.02  &  0.03  &  0.13  & 26 \\
 $\rm          [Sc/Fe] $  & -0.01  &  0.06  &  0.17  &  6 \\
 $\rm          [Ti/Fe] $  & -0.04  &  0.04  &  0.13  & 65 \\
 $\rm           [V/Fe] $  & -0.06  &  0.03  &  0.13  & 30 \\
 $\rm          [Cr/Fe] $  & -0.04  &  0.03  &  0.18  & 37 \\
 $\rm          [Mn/Fe] $  & -0.01  &  0.07  &  0.17  & 14 \\
 $\rm          [Co/Fe] $  & -0.04  &  0.02  &  0.16  & 25 \\
 $\rm          [Ni/Fe] $  & -0.04  &  0.02  &  0.15  & 41 \\
 $\rm          [Zn/Fe] $  & -0.22  &  0.13  &  0.16  &  2 \\
 $\rm           [Y/Fe] $  & -0.03  &  0.06  &  0.14  & 13 \\
 $\rm          [Zr/Fe] $  &  0.08  &  0.12  &  0.17  &  6 \\
 $\rm          [Mo/Fe] $  & -0.05  &  0.21  &  0.14  &  2 \\
 $\rm          [Ba/Fe] $  &  0.13  &  0.05  &  0.18  &  3 \\
 $\rm          [La/Fe] $  & -0.06  &  0.09  &  0.18  &  4 \\
 $\rm          [Ce/Fe] $  & -0.00  &  0.08  &  0.19  &  4 \\
 $\rm          [Nd/Fe] $  &  0.16  &  0.08  &  0.19  & 10 \\
 $\rm          [Eu/Fe] $  &  0.07  &  0.07  &  0.19  &  1 \\
\hline
\end{tabular}
\tablefoot{
$^a$ Standard deviations; 
$^b$ Median numbers of spectral lines used. Note that the available spectral lines in UVES spectra are greater or equal to these numbers. 
}
\end{table*}

\begin{table*}
\caption{Cluster information.}              
\label{tabcl}      
\centering                                      
\begin{tabular}{l c l l c c l}          
\hline\hline                        
Name & E($B-V$) & Age & R$_{\rm GC}$ & M$_{\rm TO}^a$ & [Fe/H]$^b$ & Reference\\    
 & (mag)   &  (Gyr) & (kpc)  &  (M$_{\odot}$) & (dex) &\\
\hline                                   
NGC 6802 &  0.84   &  0.90$\pm$0.1  &   7.09$^c$   & 1.9 & $+0.10\pm0.02$& This work\\
Be 81    &  0.85   &  0.98$\pm$0.1   &   5.45  & 2.2 & $+0.23\pm0.06$& \citet{Magrini2015}\\
NGC 6705 &  0.43   &  0.30$\pm$0.05  &   6.3   &3.2  & $+0.08\pm0.06$&\citet{CantatGaudin2014}\\
Tr 20    &  0.33   &  1.50$\pm$0.15  &   6.88  &1.8  & $+0.12\pm0.05$&\citet{Donati2014}\\
NGC 4815 &  0.72   &  0.57$\pm$0.07  &   6.9   &2.5  & $+0.00\pm0.04$&\citet{Friel2014}\\
\hline                                             
\end{tabular}
\tablefoot{E($B-V$), Age, and R$_{\rm GC}$ of Be81, NGC 6705, Tr20, and NGC 4815 come from \citet[][and reference to the original papers therein]{Magrini2015}. $^a$ From \citet{Smiljanic2016}. $^b$ iDR4 measurements from UVES observed cluster giants. $^c$ Assuming R$_{\odot}=8$ kpc. See text for detailed descriptions.}
\end{table*}


\end{document}